\input harvmac
\input epsf
\input amssym

\noblackbox

\def\1{{\ds 1}}

\def\CP {{\cal P }}
\def\CL {{\cal L}}

\def\CO {{\cal O}}


\def\cO{{\cal O}}
\def\CO {{\cal O}}

\def\cP{{\cal P}}
\def\CP {{\cal P }}

\def\Tr{{\rm Tr}}

\def\mathbb#1{#1}
\def\mathrm#1{\hbox{#1}}
\def\text#1{{\hbox{#1}}}
\def\mathbf#1{\hbox{\bf #1}}
\def\frac#1#2{{#1\over #2}}
\def\emph#1{{\it #1}}

\def\lfm#1{\medskip\noindent\item{#1}}

\def\foursqr#1#2{{\vcenter{\vbox{
    \hrule height.#2pt
    \hbox{\vrule width.#2pt height#1pt \kern#1pt
    \vrule width.#2pt}
    \hrule height.#2pt
    \hrule height.#2pt
    \hbox{\vrule width.#2pt height#1pt \kern#1pt
    \vrule width.#2pt}
    \hrule height.#2pt
        \hrule height.#2pt
    \hbox{\vrule width.#2pt height#1pt \kern#1pt
    \vrule width.#2pt}
    \hrule height.#2pt
        \hrule height.#2pt
    \hbox{\vrule width.#2pt height#1pt \kern#1pt
    \vrule width.#2pt}
    \hrule height.#2pt}}}}
\def\psqr#1#2{{\vcenter{\vbox{\hrule height.#2pt
    \hbox{\vrule width.#2pt height#1pt \kern#1pt
    \vrule width.#2pt}
    \hrule height.#2pt \hrule height.#2pt
    \hbox{\vrule width.#2pt height#1pt \kern#1pt
    \vrule width.#2pt}
    \hrule height.#2pt}}}}
\def\sqr#1#2{{\vcenter{\vbox{\hrule height.#2pt
    \hbox{\vrule width.#2pt height#1pt \kern#1pt
    \vrule width.#2pt}
    \hrule height.#2pt}}}}
\def\square{\mathchoice\sqr65\sqr65\sqr{2.1}3\sqr{1.5}3}

\lref\CraigRK{
  N.~J.~Craig, D.~Green,
  ``On the Phenomenology of Strongly Coupled Hidden Sectors,''
JHEP {\bf 0909}, 113 (2009).
[arXiv:0905.4088 [hep-ph]].
}

\lref\GreenDA{
  D.~Green, Z.~Komargodski, N.~Seiberg, Y.~Tachikawa, B.~Wecht,
  ``Exactly Marginal Deformations and Global Symmetries,''
JHEP {\bf 1006}, 106 (2010).
[arXiv:1005.3546 [hep-th]].
}

\lref\MurayamaGE{
  H.~Murayama, Y.~Nomura, D.~Poland,
  ``More visible effects of the hidden sector,''
Phys.\ Rev.\  {\bf D77}, 015005 (2008).
[arXiv:0709.0775 [hep-ph]].
}

\lref\RoyNZ{
  T.~S.~Roy, M.~Schmaltz,
  ``Hidden solution to the mu/Bmu problem in gauge mediation,''
Phys.\ Rev.\  {\bf D77}, 095008 (2008).
[arXiv:0708.3593 [hep-ph]].
}

\lref\LutyYE{
  M.~A.~Luty and T.~Okui,
  ``Conformal technicolor,''
JHEP {\bf 0609}, 070 (2006).
[hep-ph/0409274].
}

\lref\RattazziPE{
  R.~Rattazzi, V.~S.~Rychkov, E.~Tonni, A.~Vichi,
  ``Bounding scalar operator dimensions in 4D CFT,''
JHEP {\bf 0812}, 031 (2008).
[arXiv:0807.0004 [hep-th]].
}

\lref\RattazziGJ{
  R.~Rattazzi, S.~Rychkov, A.~Vichi,
  ``Central Charge Bounds in 4D Conformal Field Theory,''
Phys.\ Rev.\  {\bf D83}, 046011 (2011).
[arXiv:1009.2725 [hep-th]].
}

\lref\CaraccioloBX{
  F.~Caracciolo, V.~S.~Rychkov,
  ``Rigorous Limits on the Interaction Strength in Quantum Field Theory,''
Phys.\ Rev.\  {\bf D81}, 085037 (2010).
[arXiv:0912.2726 [hep-th]].
}

\lref\RattazziYC{
  R.~Rattazzi, S.~Rychkov, A.~Vichi,
  ``Bounds in 4D Conformal Field Theories with Global Symmetry,''
J.\ Phys.\ A {\bf A44}, 035402 (2011).
[arXiv:1009.5985 [hep-th]].
}

\lref\PolandWG{
  D.~Poland, D.~Simmons-Duffin,
  ``Bounds on 4D Conformal and Superconformal Field Theories,''
JHEP {\bf 1105}, 017 (2011).
[arXiv:1009.2087 [hep-th]].
}

\lref\VichiUX{
  A.~Vichi,
  ``Improved bounds for CFT's with global symmetries,''
[arXiv:1106.4037 [hep-th]].
}

\lref\PolandEY{
  D.~Poland, D.~Simmons-Duffin, A.~Vichi,
  ``Carving Out the Space of 4D CFTs,''
[arXiv:1109.5176 [hep-th]].
}

\lref\SchmaltzQS{
  M.~Schmaltz, R.~Sundrum,
  ``Conformal Sequestering Simplified,''
JHEP {\bf 0611}, 011 (2006).
[hep-th/0608051].
}

\lref\IntriligatorJJ{
  K.~A.~Intriligator, B.~Wecht,
  ``The Exact superconformal R symmetry maximizes a,''
Nucl.\ Phys.\  {\bf B667}, 183-200 (2003).
[hep-th/0304128].
}

\lref\LeighEP{
  R.~G.~Leigh, M.~J.~Strassler,
  ``Exactly marginal operators and duality in four-dimensional N=1 supersymmetric gauge theory,''
Nucl.\ Phys.\  {\bf B447}, 95-136 (1995).
[hep-th/9503121].
}

\lref\KutasovXB{
  D.~Kutasov,
  ``Geometry On The Space Of Conformal Field Theories And Contact Terms,''
Phys.\ Lett.\ B {\bf 220}, 153 (1989)..
}

\lref\SeibergPQ{
  N.~Seiberg,
  ``Electric - magnetic duality in supersymmetric nonAbelian gauge theories,''
Nucl.\ Phys.\ B {\bf 435}, 129 (1995).
[hep-th/9411149].
}

\lref\BanksNN{
  T.~Banks and A.~Zaks,
  ``On the Phase Structure of Vector-Like Gauge Theories with Massless Fermions,''
Nucl.\ Phys.\ B {\bf 196}, 189 (1982)..
}

\lref\FitzpatrickHH{
  A.~L.~Fitzpatrick and D.~Shih,
  ``Anomalous Dimensions of Non-Chiral Operators from AdS/CFT,''
JHEP {\bf 1110}, 113 (2011).
[arXiv:1104.5013 [hep-th]].
}

\lref\DobrevQV{
  V.~K.~Dobrev and V.~B.~Petkova,
  ``All Positive Energy Unitary Irreducible Representations of Extended Conformal Supersymmetry,''
Phys.\ Lett.\ B {\bf 162}, 127 (1985).
}

\lref\MackJE{
  G.~Mack,
  ``All Unitary Ray Representations of the Conformal Group SU(2,2) with Positive Energy,''
Commun.\ Math.\ Phys.\  {\bf 55}, 1 (1977).
}

\lref\LutyYE{
  M.~A.~Luty and T.~Okui,
  ``Conformal technicolor,''
JHEP {\bf 0609}, 070 (2006).
[hep-ph/0409274].
}

\lref\HanakiXF{
  K.~Hanaki and Y.~Ookouchi,
  ``Light Gauginos and Conformal Sequestering,''
Phys.\ Rev.\ D {\bf 83}, 125010 (2011).
[arXiv:1003.5663 [hep-ph]].
}

\lref\OsbornQU{
  H.~Osborn,
  ``N=1 superconformal symmetry in four-dimensional quantum field theory,''
Annals Phys.\  {\bf 272}, 243 (1999).
[hep-th/9808041].
}

\lref\DineDV{
  M.~Dine, P.~J.~Fox, E.~Gorbatov, Y.~Shadmi, Y.~Shirman and S.~D.~Thomas,
  ``Visible effects of the hidden sector,''
Phys.\ Rev.\ D {\bf 70}, 045023 (2004).
[hep-ph/0405159].
}

\lref\CohenQC{
  A.~G.~Cohen, T.~S.~Roy and M.~Schmaltz,
  ``Hidden sector renormalization of MSSM scalar masses,''
JHEP {\bf 0702}, 027 (2007).
[hep-ph/0612100].
}

\lref\PerezNG{
  G.~Perez, T.~S.~Roy and M.~Schmaltz,
  ``Phenomenology of SUSY with scalar sequestering,''
Phys.\ Rev.\ D {\bf 79}, 095016 (2009).
[arXiv:0811.3206 [hep-ph]].
}

\lref\KaplanAC{
  D.~E.~Kaplan, G.~D.~Kribs and M.~Schmaltz,
  ``Supersymmetry breaking through transparent extra dimensions,''
Phys.\ Rev.\ D {\bf 62}, 035010 (2000).
[hep-ph/9911293].
}

\lref\ChackoMI{
  Z.~Chacko, M.~A.~Luty, A.~E.~Nelson and E.~Ponton,
  ``Gaugino mediated supersymmetry breaking,''
JHEP {\bf 0001}, 003 (2000).
[hep-ph/9911323].
}

\lref\SchmaltzGY{
  M.~Schmaltz and W.~Skiba,
  ``Minimal gaugino mediation,''
Phys.\ Rev.\ D {\bf 62}, 095005 (2000).
[hep-ph/0001172].
}

\lref\DumitrescuHA{
  T.~T.~Dumitrescu, Z.~Komargodski, N.~Seiberg and D.~Shih,
  ``General Messenger Gauge Mediation,''
JHEP {\bf 1005}, 096 (2010).
[arXiv:1003.2661 [hep-ph]].
}

\lref\NelsonSN{
  A.~E.~Nelson and M.~J.~Strassler,
  ``Suppressing flavor anarchy,''
JHEP {\bf 0009}, 030 (2000).
[hep-ph/0006251].
}

\lref\NelsonMQ{
  A.~E.~Nelson and M.~J.~Strassler,
  ``Exact results for supersymmetric renormalization and the supersymmetric flavor problem,''
JHEP {\bf 0207}, 021 (2002).
[hep-ph/0104051].
}

\lref\PolandYB{
  D.~Poland and D.~Simmons-Duffin,
  ``Superconformal Flavor Simplified,''
JHEP {\bf 1005}, 079 (2010).
[arXiv:0910.4585 [hep-ph]].
}

\lref\CraigIP{
  N.~Craig,
  ``Simple Models of Superconformal Flavor,''
[arXiv:1004.4218 [hep-ph]].
}

\lref\CraigVS{
  N.~J.~Craig and D.~R.~Green,
Phys.\ Rev.\ D {\bf 79}, 065030 (2009).
[arXiv:0808.1097 [hep-ph]].
}

\lref\BaumannYS{
  D.~Baumann and D.~Green,
  ``Desensitizing Inflation from the Planck Scale,''
JHEP {\bf 1009}, 057 (2010).
[arXiv:1004.3801 [hep-th]].
}

\lref\BaumannNU{
  D.~Baumann and D.~Green,
  ``Inflating with Baryons,''
JHEP {\bf 1104}, 071 (2011).
[arXiv:1009.3032 [hep-th]].
}

\lref\KutasovIY{
  D.~Kutasov, A.~Parnachev and D.~A.~Sahakyan,
  ``Central charges and U(1)(R) symmetries in N=1 superYang-Mills,''
JHEP {\bf 0311}, 013 (2003).
[hep-th/0308071].
}

\lref\BarnesJJ{
  E.~Barnes, K.~A.~Intriligator, B.~Wecht and J.~Wright,
  ``Evidence for the strongest version of the 4d a-theorem, via a-maximization along RG flows,''
Nucl.\ Phys.\ B {\bf 702}, 131 (2004).
[hep-th/0408156].
}

\lref\MartinZK{
  S.~P.~Martin and M.~T.~Vaughn,
  ``Two loop renormalization group equations for soft supersymmetry breaking couplings,''
Phys.\ Rev.\ D {\bf 50}, 2282 (1994), [Erratum-ibid.\ D {\bf 78}, 039903 (2008)].
[hep-ph/9311340].
}

\lref\GiudiceBP{
  G.~F.~Giudice and R.~Rattazzi,
  ``Theories with gauge mediated supersymmetry breaking,''
Phys.\ Rept.\  {\bf 322}, 419 (1999).
[hep-ph/9801271].
}

\lref\KonishiHF{
  K.~Konishi,
  ``Anomalous Supersymmetry Transformation of Some Composite Operators in SQCD,''
Phys.\ Lett.\ B {\bf 135}, 439 (1984)..
}

\lref\CachazoRY{
  F.~Cachazo, M.~R.~Douglas, N.~Seiberg and E.~Witten,
  ``Chiral rings and anomalies in supersymmetric gauge theory,''
JHEP {\bf 0212}, 071 (2002).
[hep-th/0211170].
}

\lref\NovikovUC{
 ÊV.~A.~Novikov, M.~A.~Shifman, A.~I.~Vainshtein and V.~I.~Zakharov,
 Ê``Exact Gell-Mann-Low Function of Supersymmetric Yang-Mills Theories from Instanton Calculus,''
Nucl.\ Phys.\ B {\bf 229}, 381 (1983)..
}

\def\cP{{\cal P}}

\def\cO{{\cal O}}

\def\cL{{\cal L}}
\def\Tr{\mathop{\mathrm{Tr}}}

\def\cO{{\cal O}}
\def\cS{{\cal S}}


\Title{\vbox{\baselineskip12pt }} {\vbox{\centerline{Bounds on SCFTs}\vskip6pt\centerline{ from Conformal Perturbation Theory}}}

\smallskip
\bigskip
\centerline{Daniel Green$^1$ and David Shih$^2$}
\smallskip
\bigskip
\centerline{$^1${\it School of Natural Sciences,
Institute for Advanced Study, Princeton, NJ 08540 USA}} \centerline{$^2${\it 
NHETC, Dept.\ of Physics, Rutgers University, Piscataway, NJ 08854 USA}}
\vskip 1cm

\bigskip

\noindent

The operator product expansion (OPE) in 4d (super)conformal field theory is of broad interest, for both formal and phenomenological applications. 
In this paper, we use conformal perturbation theory to study the OPE of nearly-free fields coupled to SCFTs.
Under fairly general assumptions, we show that the OPE of a chiral operator of dimension $\Delta = 1+\epsilon$ with its complex conjugate always contains an operator of dimension less than $2 \Delta$.  Our bounds apply to Banks-Zaks fixed points and their generalizations, as we illustrate using several examples.

\bigskip

\Date{March 2012}

\eject
\newsec{Introduction}

Conformal field theories (CFTs) and superconformal field theories (SCFTs) play a valuable role in our understanding of quantum field theories.  Many interacting theories are believed to flow to (S)CFTs, while many more can be understood as small deformations away from conformality. Furthermore, through the AdS/CFT correspondence, CFTs are a useful tool for understanding quantum gravity.

Combined with unitarity, the enhanced symmetry of these theories has long been known to impose significant constraints on the spectrum of operator dimensions \refs{\MackJE{,}\DobrevQV}. These ``unitarity bounds" have played a valuable role in our understanding of the phases of supersymmetric gauge theories \SeibergPQ. In addition, $a$-maximization \IntriligatorJJ~has made it possible to compute the dimensions of chiral operators in proposed IR SCFTs, and unitarity bounds can be used to test the consistency of such proposals.  

In recent years, there has been significant progress in using conformal invariance to constrain the spectrum of primaries that can appear in the operator product expansion (OPE) of scalar operators.  Specifically, given a CFT with a  scalar primary operator $X$ of dimension $\Delta_X$, the OPE takes the form
\eqn\defope{
X^\dagger (x) X(0) = \frac{1}{x^{2\Delta_X}} + \sum_i \frac{c_i}{x^{2 \Delta_X - \Delta_i} } \cS_i + \ldots  \ , }
where the $c_i$ are the OPE coefficients, $\Delta_i$ are the dimensions of the scalar primary operators ${\cal S}_i$ and $\ldots$ signifies higher spin operators and conformal descendants. Applying the OPE to the four point function of $X$, \refs{\RattazziPE\RattazziGJ \CaraccioloBX{--} \RattazziYC}~were able to place an upper bound on 
\eqn\gammaphisq{
\delta_{min}\equiv \Delta_{min}-2\Delta_X
}
where $\Delta_{min}$ denotes the smallest $\Delta_i$ that appears in the OPE.  Subsequent work has extended these results, including to supersymmetric theories \refs{\PolandWG \VichiUX{--}\PolandEY}.

In the context of supersymmetry, there has been much interest in theories with large anomalous dimensions, in part because of their many applications to model building, see e.g.\ \refs{\DineDV\RoyNZ\MurayamaGE\CraigRK\CohenQC\PerezNG\KaplanAC\ChackoMI\SchmaltzGY\CraigVS\DumitrescuHA\HanakiXF\NelsonSN\NelsonMQ\PolandYB\CraigIP\BaumannYS-\BaumannNU} (for non-SUSY motivation, see e.g.\ \refs{\RattazziPE{,}\LutyYE} and references therein).  For example, one can try to resolve the $\mu$/$B \mu$ problem by using an SCFT with $\delta_{min}>0$ \refs{\DineDV\RoyNZ\MurayamaGE-\CraigRK}.  We often imagine generating $\mu$ and $B \mu$ at the SUSY breaking scale through the F-term of a chiral operator $X$ via the effective lagrangian
\eqn\mubmu{
{\cal L} \supset \int d^4 \theta\left[ \  c_\mu \left( \frac{X^{\dagger} }{\Lambda}H_u H_d + {\rm h.c.} \right)+ c_{B\mu} \frac{X^{\dagger}X }{\Lambda^2} (H_u H_d +{\rm h.c.} ) \right] \ . }
Electroweak symmetry breaking requires $c_{B\mu}\sim c_\mu^2$ at the scale SUSY breaking.  Unfortunately, models almost invariably give rise to $c_{B\mu} \gg c_{\mu}^2$ at the scale where they are generated -- this is known as the $\mu/B\mu$ problem (see \GiudiceBP\ for a review and original references).  However, if the theory flows near a SCFT in between these scales, the ``$X^{\dagger} X$" is replaced by the operators in the OPE of $X^{\dagger}$ and $X$.  If all\foot{One may weaken this requirement to allow conserved current multiplets $J$ (i.e.\ real multiplets of dimension two) to appear in the OPE \SchmaltzQS.  This is viable because $D^2 J = 0$, by definition, and therefore does not contribute to $B \mu$.  However, one cannot spontaneously break SUSY in an SCFT. Therefore, the SCFT must be deformed by relevant operators in order to break conformal invariance.  For this to be a valid solution to the $\mu-B\mu$ problem, one must show that the deformation can be chosen such that the current remains conserved at all energies.  We will ignore this possibility, as it requires knowledge of the field theory that cannot be determined from the behavior near the fixed point alone.} such operators have dimension $\Delta_i > 2 \Delta_X$, then RG flow will suppress $c_{B\mu}$ down to acceptable levels in the IR.

Although the bounds in \refs{\PolandWG \VichiUX{--}\PolandEY}\ allow for positive $\delta_{min}$, to date there is no existence proof showing that this is possible. In \FitzpatrickHH, SCFTs with weakly coupled AdS duals were studied, and it was shown that $\delta_{min}$ could have either sign as far as the 5d effective gravity theory was concerned. If these models could be UV completed into full string compactifications, then they would furnish existence proofs of positive $\delta_{min}$. Whether this is possible remains an interesting open question.

In this paper, we will take a complementary approach, and investigate the quantity $\delta_{min}$ in the context of SCFTs in which $X$ is a chiral primary operator with dimension $\Delta_X=1+\epsilon$, with $\epsilon\ll 1$. When $\Delta_X =1$, the unitarity bounds require that $X$ is a free field and its OPE is trivially determined. One might suspect that for $\Delta_X=1+\epsilon$, the bounds on operators in the OPE are accessible perturbatively in $\epsilon$.  We will develop a general formalism for computing $\delta_{min}$ based on conformal perturbation theory, with the assumption that $X$ acquires its anomalous dimension solely through a coupling in the superpotential to a chiral operator $\CO$ of  dimension $\Delta_\CO=2-\epsilon$.  Under such circumstances, the theory will flow to a fixed point where $X$ has the required dimension, and we will prove that the OPE of $X$ and $X^{\dagger}$ contains an operator of dimension at most $\Delta = 2\Delta_X + |\cO(\epsilon^2)|$.  We will further show that $\delta_{\rm min} \sim  - \epsilon^2/ \nu_i < 0$ when the original SCFT contains real operators $L_i$ of dimension $\Delta_i = 2 + \nu_i$ with $\nu_i \ll 1$. As an application of our formalism, we will prove that in fully perturbative examples, such as generalized Banks-Zaks fixed points, one always finds $\delta_{min}<0$.

While these assumptions may seem restrictive, it is plausible that this setup actually applies to all examples of SCFTs with a chiral operator with dimension close to one.  In the limit $\epsilon \to 0$, the operator $X$ must become free and will decouple from any other fields.  Furthermore, $X$ must be gauge invariant.  Provided there is some Lagrangian description of $X$, this suggests the above description should hold in some duality frame.

The organization of the paper is as follows:  In Section 2, we will explain how the bounds on scalar operators in the OPE of chiral primaries in an SCFT can be understood from conformal perturbation theory.  The presentation in this section will be mostly general, making only a few simplifying assumptions about the form of the SCFT.  In Section 3, we will compare these results with concrete examples derived from Banks-Zaks-like SCFTs.  The results in this section agree with the results of conformal perturbation theory but are derived in a more conventional language.  We will summarize our results and mention some interesting open problems in Section 4. Appendix A contains the details of the conformal perturbation theory calculations. Appendix B addresses complications arising from the presence of global symmetries.  Appendix C contains further details regarding the matching between the results of Section 2 and 3.

\newsec{Scaling Behavior of Nearly Free Fields}
\subsec{Setup}

Consider a 4d interacting SCFT $\cP_1$ which contains a chiral operator $\CO$ of dimension $\Delta_\CO = 2-\epsilon$.   The OPE of $\CO$ and $\CO^\dagger$ is given by:
\eqn\Oope{
\cO(x)^\dagger \cO(0) =  |x|^{-2 \Delta_{\cO}} + \sum_i |x|^{\Delta_{i} - 2 \Delta_{\cO} } c_i L_i + \ldots\ ,}  
where $L_i$ are real scalar multiplets with dimension $\Delta_i  = 2+\nu_i$, $\nu_i\ge 0$, and $\dots$ denotes descendants and operators with higher spin.  All primary scaling operators in this paper will be ``CFT-canonically normalized" as in \Oope, i.e.\ with the coefficient of the leading singularity set to one.

Now consider adding to $\cP_1$ a free chiral superfield $X$, and the deformation
\eqn\deformation{
\cL = \cL_{\cP_1}+ {1\over4\pi^2}\int d^4 \theta\, X^{\dagger} X + \left( \int d^2 \theta \,\frac{\lambda }{2\pi} \Lambda^{\epsilon} \cO X + {\rm h. c.} \right) \ ,}
where we have introduced the Wilsonian cutoff  scale $\Lambda$ in order to make $\lambda$ dimensionless.  Note that $X$ is not a canonically normalized field; this is to obey the more natural CFT normalization described in the previous paragraph. When $0< \epsilon \ll 1$, this theory will flow to a new fixed point, $\cP_2$, with $\lambda \sim \cO(\sqrt{\epsilon})$.  At the new fixed point, we would like to know what is the lowest dimension scalar operator that appears in the OPE of $X$ and $X^{\dagger}$. Candidate operators include $X^\dagger X$ and the $L_i$.  In general, these are not scaling operators at the new fixed point $\cP_2$, but will instead mix with one another. As we will see in the following subsections, when the dimensions of $L_i$ are approximately 2, this mixing is the dominant contribution to $\delta_{\rm min}$.

We will focus on the case where conserved currents are absent in the $X^\dagger(x)X(0)$ OPE. In the free theory, $X^{\dagger} X$ is the conserved current associated with the symmetry $U(1)_X$.  This current is broken by \deformation\ but, in general, new conserved currents could arise from a combination of $U(1)$ currents broken by the superpotential interaction, i.e.\ a combination of $U(1)_X$ and $U(1)$'s from $\CP_1$ under which $\CO$ is charged.  The unitarity bounds ensure that any such current would have dimension two, and therefore would trivially satisfy $\Delta_{\min} < 2 \Delta_X$ for $\Delta_X>1$.

To simplify the discussion in this section, we will assume that $\CO$ is a singlet under all the global symmetries of $\cP_1$, i.e.\ the $\nu_i$ are strictly positive in \Oope. This ensures that the deformation \deformation\ breaks the $U(1)_X$ symmetry completely, and then  absence of conserved currents is trivially guaranteed.  In appendix B, we will generalize to the (considerably more complicated) case of charged $\CO$. With some mild assumptions, all the results of this section carry over to charged $\CO$, provided that conserved currents remain absent in the $X^\dagger X$ OPE.

\subsec{RG flow}

Under the assumption that $\epsilon \ll 1$, we can understand the RG flow induced by \deformation\ using conformal perturbation theory.  Without loss of generality, $\lambda$ will be real, as its phase can be removed by a field redefinition of $X$.  As above, $\Lambda$ will denote the sliding Wilsonian cutoff scale.

Once we introduce the deformation by a single operator, as in \deformation, other operators will also contribute to the RG flow at higher orders in $\lambda$.  Therefore, the RG is actually described by the flow in a larger space of couplings.  We will include these additional operators in the Lagrangian as
\eqn\rgaction{
\cL = \cL_{\cP_1} + \int d^4 \theta\, \left( {1\over4\pi^2}Z_X X^{\dagger} X +  \sum_i y_i \Lambda^{-\nu_i} L_i + \ldots  \right)+  \left( \int d^2 \theta\, \frac{\lambda}{2\pi} \Lambda^{\epsilon} \cO X  + {\rm h. c.} \right) \ .}
Here $Z_X$ is the wavefunction renormalization of $X$ and $y_i$ are coupling constants.  The $\ldots$ denote operators that that do not appear in the $\CO^\dagger\CO$ OPE. The counterterms for these can only arise at higher orders in $\lambda$, and so they will not contribute to our analysis in this paper. We have also dropped operators of the form $X^{\dagger} X L_i$ as they contribute at the same order as $L_i$ of dimension $\Delta_i > 4$ and will be negligible in the regimes of interest.  We have included explicit $\Lambda$ dependence to make all the couplings dimensionless.  By a non-renormalization theorem \GreenDA, we have chosen a holomorphic scheme such that no additional operators appear in the superpotential.

If we work in a holomorphic scheme, there is no renormalization of $\lambda$ and its beta function $\beta_\lambda\equiv {d \lambda\over d\log\Lambda}$ is given purely by the classical contribution, $\beta_\lambda = - \epsilon \lambda$.   
 In addition, there is wavefunction normalization, namely $Z_X = \Big( \frac{\Lambda_0}{\Lambda} \Big)^{2 \gamma_X (\lambda, y_i)} = 1 + 2 \gamma_X (\lambda, y_i) \log(\Lambda_0 /\Lambda) + \ldots$ (where $\Lambda_0$ is some arbitrary reference scale).  In this scheme, as we flow to the IR ($\Lambda \to 0$), $\lambda$ and $Z_X$ will run to infinity.  To see that we reach a fixed point, we redefine $X$ and $\lambda$ to make the former (nearly) canonical
\eqn\redefineXlambda{
 X \to  X^{phys} = Z_X^{1/2}  X,\qquad \lambda \to \lambda^{phys}= Z_X^{-1/2}\lambda \ .
} 
After the redefinition, the beta function of $\lambda$ (henceforth dropping the ``phys" superscripts) becomes
\eqn\betalambdanew{
\beta_{\lambda} = - \epsilon \lambda + \lambda \gamma_X (\lambda, y_i) \ .
} 
Regardless of the form of $\gamma_X (\lambda, y_i)$, we see that the fixed point will arise at $\gamma_X (\lambda_\star, y_{i \, \star}) = \epsilon$.  As expected, the dimension of $X$ at $\cP_2$ is $\Delta_X = 1 + \gamma_X (\lambda_\star, y_{i \, \star}) = 1  + \epsilon$.

In order to the determine the dimensions of operators at $\cP_2$, we will need to know the functional form of $\gamma_X$.  Expanding $\gamma_X$ in powers of $\lambda$ and $y_i$, we find 
\eqn\gammaform{
\gamma_X = \pi^2  \lambda^2 - 4\pi^4\sum_i \CI(\nu_i,\epsilon) c_i y_i \lambda^2+\dots \ .
}
We have relegated the detailed computations to appendix A. Here $c_i$ is the OPE coefficient of $\cO \cO^{\dagger} \to L_i$ in \Oope.  $\CI$ is a function of $\nu_i$ and $\epsilon$ which is smooth everywhere, and which we have normalized so that:
\eqn\CIlimit{
\lim_{\nu_i,\epsilon\to 0}\CI(\nu_i,\epsilon) = 1 \ .
}
The $\dots$ in \gammaform\ includes important corrections at $\CO(\lambda^4)$. Ideally we would also compute these corrections for completeness. However, the calculation is third order in conformal perturbation theory and beyond the scope of this work.

Finally, we also need the leading-order beta functions for the couplings $y_i$.  There is a classical contribution to the beta functions from the explicit scale dependence, $\beta_{y_i} = \nu_i y_i + \ldots$.  The leading order correction in $\lambda$ was already determined using conformal perturbation theory in \CraigRK; we have redone it in appendix A with the notation and conventions of this paper for completeness sake. As a result of this calculation, one finds 
\eqn\betay{
\beta_{y_i} = \nu_i y_i - \frac{1}{2} c_{i}  \lambda^2 + \ldots \ ,}
where $c_i$ is again the same OPE coefficient as appears in \Oope\ and \gammaform. We see that $y_i$ will flow to a fixed point $y_{i \star} \approx \frac{1}{2} \frac{c_i}{\nu_i}  \lambda_\star^2$.  

For the theory to be under control and conformal perturbation theory to be valid, we obviously need $y_{i\star} \ll 1$. This places a restriction on the range of $\nu_i$ relative to the other parameters. In general, we need 
\eqn\nuilowerlim{
\nu_i\gg c_i\epsilon \ .
} 
If $c_i\sim\CO(1)$, then $\nu_i$ must be parametrically larger than $\epsilon$. If, on the other hand, $c_i\ll 1$ (as in the perturbative examples we will study in section 3), then $\nu_i$ can be much smaller. Of course, theories where $\nu_i$ violates \nuilowerlim\ are perfectly valid, they just cannot be described using conformal perturbation theory around $y_i=0$. Instead, it makes more sense to treat the corresponding operators $L_i$ as nearly-conserved currents. In appendix B, we outline how to treat exactly conserved currents; we expect that nearly-conserved currents behave in a similar way. We will leave a complete analysis to future work.

In summary, the RG flow of \rgaction~is described by the system of beta functions
\eqn\betas{\eqalign{
\beta_{\lambda} &= - \epsilon \lambda + \lambda \left(\pi^2 \lambda^2 -4\pi^4 \sum_i  c_i\CI(\nu_i,\epsilon)  y_i \lambda^2 + \ldots\right) \cr
 \beta_{y_i} & = \nu_i y_i - \frac{1}{2} c_{i} \lambda ^2 + \ldots \ . }}
The fixed point $\cP_2$ is attractive and occurs when the couplings are
\eqn\fixedcouplings{ \eqalign{
& y_{i \star} = \frac{c_i }{2 \pi^2 \nu_i} \epsilon + \ldots \ , \cr
 &\lambda_\star^2 = \frac{\epsilon}{ \pi^2} + {2\over \pi^2}   \sum_i c_i^2 \CI(\nu_i,\epsilon)  { \epsilon^2\over \nu_i} + \ldots \ .}}
 This is sufficient information to determine the dimensions of the non-chiral, scalar operators at $\cP_2$, which we will turn to in the next subsection.

\subsec{Operator dimensions}

Given the beta functions that describe an RG flow, it is straightforward to determine the set of scaling operators and their dimensions that describe the infinitesimal deformations away from a fixed point.  Specifically, if we deform any fixed point by $\int d^4 x\, \kappa\, \Lambda^{4 - \Delta} \Sigma$ where $\Sigma$ is some operator of dimension $\Delta$, the beta function for the coupling $\kappa$ must take the form $\beta_{\kappa} = (\Delta- 4)\kappa+ \cO(\kappa^2)$.  Reversing this logic, if we are given the beta functions for some couplings $\kappa_i$ near a fixed point, we can determine the anomalous dimensions of operators at that fixed point by Taylor expanding and diagonalizing the beta functions around it,
\eqn\dimbeta{
\beta_{\kappa_i} =  \frac{\partial \beta_{\kappa_i}}{\partial \kappa_j} \Bigg|_{\kappa = \kappa_{\star} } (\kappa_j-\kappa_{j\star}) + \cO((\kappa-\kappa_{\star}) ^2)\ .}
The eigenvectors and eigenvalues of $\partial_{\kappa_j} \beta_{\kappa_i}$ determine the scaling operators and anomalous dimensions, respectively.

Applying this to the beta functions \betas\ at the fixed point \fixedcouplings, we find the matrix of anomalous dimensions at $\CP_2$:\foot{To keep the presentation from becoming too cluttered, we will simply drop the higher-order terms in conformal perturbation theory in the following formulas. We will discuss some of these corrections shortly. In the next subsection, we will analyze their effects more systematically. }
\eqn\scalematrix{
\Gamma \equiv  \partial_{\{y_j , \lambda\}} \beta_{\{y_i , \lambda\}} |_{y_i =y_{i \star} , \lambda = \lambda_\star}  = \pmatrix{ {\nu_i \delta_{i j} } & { -c_i \lambda_\star} \cr {-4\pi^4 c_i \CI(\nu_i,\epsilon)  \lambda_\star^3} & {2 \epsilon} } \ .}
As discussed in the introduction, we are interested in the anomalous dimensions relative to $2\gamma_X=2\epsilon$. It is convenient to introduce the shifted matrix
\eqn\scalematrixii{
\Delta\Gamma \equiv  \Gamma-2\epsilon\,{\bf 1} = \pmatrix{ {(\nu_i-2\epsilon) \delta_{i j} } & { -c_i \lambda_\star} \cr {-4\pi^4 c_i \CI(\nu_i,\epsilon)  \lambda_\star^3} & 0 } \ .
}
The eigenvalues $\delta$  of this matrix satisfy the characteristic equation
\eqn\eigen{
 \left( \delta + 4 \epsilon^2 \sum_j  {c_j^2  \CI(\nu_j,\epsilon) \over \nu_j - 2\epsilon-\delta} \right) \times \prod_i (\nu_i - 2\epsilon-\delta ) = 0 \ .}
where we have used \fixedcouplings\ to replace $\lambda_\star$ with $\sqrt{\epsilon}/\pi$ (again, dropping higher order corrections for now). The minimum eigenvalue $\delta_{min} \equiv \Delta_{\rm min} - 2 \Delta_X$ is the quantity of interest, introduced in \gammaphisq. In order to understand the behavior of $\delta_{min}$, it will now be useful to distinguish different scenarios for the $\nu_i$:

\lfm{\bf Scenario 1:} $\nu_i \ll 1$ for some $i$. This scenario applies to all fully perturbative SCFTs (generalized Banks-Zaks models), because in these theories the kinetic terms are always approximately dimension 2. In the $\nu_i\ll 1$ regime, two things happen. First, $\CI(\nu_i,\epsilon)\to 1$ as shown in appendix A. Then $\Delta\Gamma$ is related by a similarity transform to a symmetric matrix
\eqn\Gammaprime{
\Delta\Gamma_{sym} = \pmatrix{(\nu_i-2\epsilon)\delta_{ij} & -2 c_i \epsilon \cr -2c_i \epsilon & 0}
}
The minimum eigenvalue of any symmetric matrix is always smaller than all the diagonal elements, so we conclude that 
\eqn\deltaminneg{
\delta_{min}<0
} 
in this regime. The second thing that happens in this regime is that the mixing with $X^\dagger X$ is enhanced, making its contribution to the anomalous dimension reliable compared to the $\CO(\epsilon^2)$ from higher order in conformal perturbation theory. The situation is clearest if $\epsilon\ll\nu_i\ll 1$, where we find from \eigen
\eqn\eigensolapproxii{
\delta_{min} \approx -{4\epsilon^2 c_i^2\over \nu_i} \ .
}
This is indeed enhanced relative to the $\CO(\epsilon^2)$ corrections to the beta functions which we have not computed. If on the other hand $\epsilon\sim\nu_i\ll 1$, then \eigensolapproxii\ no longer applies and we have to do degenerate perturbation theory instead. But, in this case, it is easy to see that the enhancement is strictly larger -- instead of \eigensolapproxii, $\delta_{min}<0$ scales as $\CO(\epsilon)$. Finally, if $\nu_i\ll\epsilon$, then there is already an operator $L_i$ in the OPE of $X^\dagger X$ with much smaller anomalous dimension than $2\epsilon$, so $\delta_{min}<0$ trivially.\foot{Technically, this assumes that the corrections to $\nu_i$ from mixing are parametrically smaller than $\nu_i$ itself. If instead quantum corrections to $\nu_i$ are large, then conformal perturbation theory is clearly breaking down. We will discuss this further in the next subsection.}

\medskip

\lfm{\bf Scenario 2:} $\nu_i \sim 1$ for all $i$. Theories in this class necessarily contain some strong dynamics. When all the operators $L_i$ have $\CO(1)$ anomalous dimensions, the mixing with $X^\dagger X$ is not enhanced, and we find 
\eqn\eigensolapprox{
\delta_{min}\approx -4\epsilon^2\sum_j {c_j^2 \CI(\nu_j,\epsilon)\over \nu_j}
}
from \eigen. In this case, we cannot determine the sign of $\delta_{min}$. Even if we knew the sign of $\CI(\nu_j,\epsilon)$, it turns out higher order corrections in conformal perturbation theory are equally important to the contribution \eigensolapprox.  In particular, $\delta_{min}$ is sensitive to $\CO(\lambda^4)$ terms in $\beta_\lambda$ which we have not computed (the $\dots$ in \betas).   Nevertheless, we see that the lowest dimension operator in the $X^{\dagger}X$ OPE has dimension $\Delta_{min} = 2\Delta_X + \delta_{min} = 2 + 2\epsilon+\CO(\epsilon^2)$ to leading order.  This is consistent with the bounds of \PolandEY, who found that $\Delta_{min} \leq 2 + 2\epsilon + 2.683 \epsilon^2$. It would be very interesting to see whether this bound is saturated after including the $\CO(\lambda^4)$ contributions. If this bound is the strongest possible one, then it may be possible to compute the numerical coefficient at $\CO(\epsilon^2)$ directly from conformal perturbation theory.

\bigskip

To summarize: in this section, we found that the lowest dimension operator that can appear in the OPE of $X$ and $X^{\dagger}$ has dimension that is at most $\Delta_{min} = 2 \Delta_X + |\cO(\epsilon^2)|$. When $\nu_i \ll 1$ for some $i$, as is the case for all fully perturbative SCFTs, we found that $\Delta_{min} - 2 \Delta_X \sim - \epsilon^2 / \nu_i < 0$.  The bounds when $\nu_i \ll 1$ are stronger than those found in \PolandEY.  It would be interesting to determine all $\cO(\epsilon^2)$ corrections to see if this is true in general.

\subsec{Taking into account higher-order corrections}

In the previous subsections, we determined the fixed point and the matrix of anomalous dimensions by neglecting higher order corrections in conformal perturbation theory. Now let us discuss the conditions under which this is valid.

One basic assumption which we have been making implicitly so far is that there are no large hierarchies in the OPE coefficients. Without this assumption, leading-order conformal perturbation theory is not necessarily valid. For instance, if the coefficient of $y\lambda^2$ in \gammaform\ is anomalously small, then higher orders in conformal perturbation theory involving unsuppressed OPE coefficients could dominate. 

In other words, we are assuming that parametrically, \gammaform\ and \betay\ are really
\eqn\gammabetaN{\eqalign{
 & \gamma_X =\pi^2 \lambda^2 +\kappa y_i \left(-4\pi^4\CI(\nu_i,\epsilon)\tilde c_i\lambda^2+\dots\right) + \kappa^2 \lambda^4(\ldots)\cr
 & \beta_{y_i} = \nu_i y_i + \kappa\left(-{1\over2}\tilde c_i\lambda^2 + \dots\right)\cr
}}
Here $\tilde c_i$ and all the coefficients in $\dots$ are assumed to be $\CO(1)$. $\kappa$ can either be (at most) $\CO(1)$ or parametrically small. In the BZ models of Section 3, $\kappa\sim 1/N$. 

Having clarified this point, let us now systematically consider higher-order corrections to $\gamma_X$ and $\beta_{y_i}$, the $\dots$ in \gammaform\ and \betay. Corrections to $\gamma_X$ must be proportional to $\lambda^2$; combining this with \gammabetaN, we conclude that they are always subleading for determining the fixed point couplings. As discussed in the previous subsection, they are also subleading for determining the anomalous dimensions, provided that there are $L_i$ operators with dimensions close to two. All higher-order corrections to $\beta_{y_i}$ proportional to $\lambda^2$ are also subleading for determining the fixed point couplings and the anomalous conditions, in general. 

This leaves corrections to $\beta_{y_i}$ which are $\CO(\lambda^0)$, i.e.\ which involve the $y$'s only. Since these should be compared with the classical term $\nu_i y_i$, they can conceivably be important when $\nu_i\ll 1$. For instance, consider an $\CO(y^2)$ correction to $\beta_{y_i}$:\foot{These corrections seemingly could lead to new fixed points with $y\sim \nu$. However, it is easy to check that such fixed points are repulsive, not attractive. Around the attractive point we considered above, $\CO(y^2)$ corrections are clearly subleading as far as determining the value of $y_*$ is concerned.}
\eqn\deltabetay{
\delta\beta_{y_i} = {1\over2}c_{ijk}y_j y_k
}
These will affect the matrix of anomalous dimensions as $\delta\Gamma_{ij} = c_{ijk}y_k\sim {c_{ijk} c_k \epsilon\over \nu_k}$. So if $\nu_i$ is too small, even if it satisfies \nuilowerlim, this perturbation to the matrix of anomalous dimensions could be larger than the classical dimension.  For this paper, we will simply {\it assume} that these higher order terms are subleading. Nevertheless, it is plausible that this assumption could be promoted to a consequence of $y_{i\star}\ll 1$.  Because the two-point function $\langle (Q^2 L_i) (\bar Q^2 L_i) \rangle \propto \nu_i$, we must have $\langle (Q^2 L_i) (Q^2 L_j) \tilde \CO \rangle \propto \nu_i \nu_j$ in order for the $ ( Q^2 L_{i,j} )\tilde \CO$ OPE coefficient to be finite as $\nu_i \to 0$.  The resulting $\nu_{i} \nu_{j}$ suppression of the $(Q^2 L_i) (\bar Q^2 L_j)$ OPE  ensures that $\delta \Gamma_{ij}$ is negligible when $y_{i,j} \ll 1$.  However, a careful derivation valid to all orders in perturbation theory is beyond the scope of this work.

More generally, as mentioned already in the previous subsection, in order for our analysis in this paper to be valid, we must avoid extremely small values of $\nu_i$. The classical approximation of $\beta_{y_i}=\nu_i y_i+\dots$ should be a reliable starting point for conformal perturbation theory.  For this to be true, $\nu_i$ must be parametrically larger than any of the corrections to the dimension of $L_i$ coming from conformal perturbation theory. In the BZ models with $\kappa\sim 1/N$ being the smallest parameter, this is automatically guaranteed. 

Overall, the complication presented in this general analysis is that, a priori, all the OPE coefficients and dimensions appear to be independent parameters.  If one of these numbers is anomalously small or large, it could invalidate our leading order calculation.  In practice, a given theory (or class of theories) is controlled by a much smaller set of parameters.  Using these parameters, it is easier to establish the regime of validity of conformal perturbation theory.

\newsec{Banks-Zaks Fixed Points}

In the previous section we presented general results on the spectrum of operators in the OPEs of a nearly free, chiral superfield $X$.  Now, we will illustrate these general results with broad class of simple, calculable examples, namely supersymmetric Banks-Zaks \refs{\BanksNN,\SeibergPQ}~(BZ) fixed points and their generalizations. We will examine the dimensions of operators at BZ fixed points from a more conventional perspective\foot{In particular, we will take formulas and conventions for perturbative beta functions and anomalous dimensions directly from  \MartinZK.} and show that the results agree with the analysis using conformal perturbation theory.  

Before we start with a detailed analysis, let us make a few general remarks about perturbative SCFTs.  In four dimensions, gauge fields are required in order to produce a non-trivial, perturbative fixed point.  This requirement has several important consequences: (1) Charged matter cannot produce chiral operators of dimension near one because the matter fields are not gauge invariant.  (2) To produce an operator of approximate dimension one, we must introduce a singlet, $X$, under the gauge group; therefore, the only marginal or relevant interactions are through operators in the superpotential. (3)  The kinetic terms for the matter fields are approximately dimension two and will appear in the OPE of $X$ and $X^{\dagger}$.  (4)  Therefore, all the models of this type are described by ``scenario 1" of the previous section and we must have $\delta_{min}<0$ on general grounds.  The two examples in this section will serve as illustrations of these basic observations.

\subsec{Simple Banks-Zaks fixed points}

The simplest supersymmetric Banks-Zaks fixed points arise in ${\cal N}=1$, $SU(N_c)$ gauge theories with $N_f = {3N_c/ (1+\epsilon)}$ flavors \SeibergPQ~(we assume $N_f$, $N_c\to\infty$ throughout).  The only coupling in these theories is the gauge coupling; since we are at large $N$, we will use instead the 't Hooft coupling
\eqn\thooftsimple{
\hat g \equiv {N_c g^2\over 8\pi^2} \ .
}
Its RG flow is described (in a suitable scheme) by the NSVZ beta function \NovikovUC:
\eqn\nsvz{\eqalign{
 \beta_{g} & = -\frac{g^3}{16 \pi^2} \frac{ \Big( 3 N_c -  N_f(1- 2\gamma_Q) \Big)}{1 - N_c g^2 / 8 \pi^2} 
\cr & \to\quad  \beta_{\hat g} =  -{6\over 1+\epsilon}{\hat g^2\over 1-\hat g}\left({\epsilon\over 2}+\gamma_Q\right) \equiv -6\hat g^2 f(\hat g)\left({\epsilon\over 2}+\gamma_Q\right)
}}
where we have defined $f(\hat g)$ so that $f(\hat g)=1+\dots$, and
\eqn\gammaQ{
\gamma_Q =\Big( -{1\over2}\hat g + \CO(\hat g^2)\Big) + {1\over N_c^2}\Big({1\over2}\hat g + \CO(\hat g^2)\Big) + \CO\Big({1\over N_c^4}\Big)
}
is the anomalous dimension of the fundamental flavors. We have written the anomalous dimension in the form \gammaQ\ to emphasize the nature of the double expansion in $1/N_c$ and $\hat g$. The Banks-Zaks fixed point occurs at $\beta_{\hat g}=0$, which according to \nsvz\ is equivalent to 
\eqn\gammaQfp{
\gamma_Q=-{1\over2}\epsilon \ ,
}
or in terms of the 't Hooft coupling,
\eqn\ghatfp{
\hat g_{*,0} =\Big( \epsilon + \CO(\epsilon^2)\Big) + {1\over N_c^2} \Big( \epsilon + \CO(\epsilon^2)\Big) +\CO\Big( {1\over N_c^4}\Big) \ .
} 
Here the 0 subscript on $\hat g_*$ is to emphasize that this is the fixed point in the undeformed BZ theory. Shortly we will deform the theory by coupling it to $X$, and that will change the value of $\hat g_*$. 

The BZ fixed point will play the role of $\cP_1$ in Section 2.  We would now like to identify some linear combination of the gauge invariant operators $Q_f\widetilde Q_{\tilde g}$ with the operator $\CO$ in section 2. According to \gammaQfp, these have dimension close to 2:
\eqn\QQdim{
\Delta(Q_f\widetilde Q_{\tilde g})  = 2-\epsilon \ .
}
Note that since $Q$ and $\tilde Q$ are not gauge invariant operators, there are no chiral primaries with dimension near 1. This is a general feature of all BZ-type theories.

As discussed in Section 2, the principal requirement on the operator $\CO$ (other than having dimension close to 2) is that after deforming by $\int d^2\theta\,\lambda X \CO$, the $U(1)_X$ symmetry is completely broken, so that no conserved currents appear in the $X^\dagger X$ OPE.\foot{Note that in Section 2, we assumed for simplicity that $\CO$ was a singlet under the global symmetries of the SCFT, whereas here $Q_f\widetilde Q_{\tilde g}$ transforms as $(\square,\square)$ under the $SU(N_f)_L\times SU(N_f)_R$ global symmetry.  
This complication is dealt with in Appendix B -- the upshot being that in generalized BZ theories, provided that no conserved current appears in the $X^\dagger X$ OPE, the generators that are broken by the deformation do not mix with $X^{\dagger}X$ and $L_i$ at the order we study and decouple from the problem. This is doubly guaranteed in BZ theories, by the fact that the $L_i$ are singlets with respect to the global symmetry, and by the fact that we are at large $N$.} Modulo $SU(N_f)_L\times SU(N_f)_R$ rotations, this condition is uniquely satisfied with the choice
\eqn\OdefsBZ{
\CO = a\, \delta^{f\tilde f}Q_f \tilde Q_{\tilde f} \ .
}
The normalization $a = {4\pi^2\over \sqrt{N_f N_c}}(1 + \cO(g^2))$ is chosen so that $\CO$ is CFT-canonically normalized (with $Q$, $\tilde Q$ canonically normalized).   The deformation breaks the flavor group $U(1)_X \times SU(N_f) \times SU(N_f) \to SU(N_f)_{\rm diag}$, so there is no unbroken $U(1)$ which can appear in the OPE of $X$ and $X^\dagger$.

To make contact with Section 2, we must also determine the OPE coefficients and dimensions at $\cP_1$.  Given \OdefsBZ,  the $\CO^\dagger\CO$ OPE is
\eqn\opeQQgen{
\CO(x)^\dagger \CO(0) \sim  |x|^{-2\Delta_\CO}+ c  |x|^{\Delta_L-2\Delta_\CO} L(0) + \dots \ .
}
Here $c$ is the OPE coefficient, and $L$ is given by
\eqn\LdefsBZ{
L =  b\, {\rm Tr}\,(\widetilde Q^\dagger\widetilde Q+Q^\dagger Q).
}
To leading order in perturbation theory (i.e.\ with free-field contractions), the coefficients $b$ and $c$ are given by
\eqn\abclo{
b =  {4\pi^2\over\sqrt{2N_fN_c}}, \quad c =\sqrt{2\over N_f N_c} \ .
}
The additional scalar operators in \opeQQgen\ can be safely ignored. Any $L_i$ must be a $SU(N_f)_{\rm diag}\times U(1)_R$ singlet and be invariant under $Q\leftrightarrow \tilde Q$, and \LdefsBZ\ is the unique such operator with approximate dimension 2. Note that the OPE coefficient is $1/N$ suppressed. This is a general feature of BZ-type theories and is important for the decoupling of global symmetry currents explicitly broken by the deformation of $\CP_1$.  We also note that $L$ is a singlet under the full $SU(N_f) \times SU(N_f)$ which leads to a further decoupling of these currents.

 We see from \LdefsBZ\ that $L$ is nothing but the anomalous Konishi current of the SQCD theory, which satisfies the anomaly equation $\bar D^2 L \propto {\rm Tr} W_{\alpha} W^{\alpha}$ \KonishiHF.  Therefore the anomalous dimensions of $L$ and $\Tr\,W_\alpha^2$ are the same. The latter can be derived by expanding the beta function for the gauge coupling around the fixed point (see the related discussion at the beginning of Section 2.3). We conclude that
\eqn\deltaW{\eqalign{
\nu_L = \partial_{\hat g} \beta_{\hat g} |_{\hat g=\hat g_{*,0}} &= -6\hat g_{*,0}^2 f(\hat g_{*,0})\partial_{\hat g}\gamma_Q(\hat g_{*,0}) \cr
&= \Big( 3\epsilon^2+\CO(\epsilon^3)\Big) +{1\over N_c^2}\Big( 3\epsilon^2+\CO(\epsilon^3)\Big) + \CO\Big({1\over N_c^4}\Big) \ ,
}}
where $f(\hat g)$ was defined in \nsvz, and we have used \gammaQ\ and \ghatfp\ in the last equality.  The anomalous dimension of $L$ is $\CO(\epsilon^2)$.  

Finally, let us deform the theory by 
\eqn\deltaWsBZ{
W = {\lambda\over2\pi} X \CO = {\lambda\, a\over 2\pi} X \delta^{f\tilde f}Q_f\tilde Q_{\tilde f}
}
and work out the matrix of anomalous dimensions in this simple example. Here $a$ was defined below \OdefsBZ. We will first compute the anomalous dimensions using a more conventional perturbative approach. Then we will compare this with the general conformal perturbation theory of Section 2. 

In the deformed theory, the anomalous dimensions of $X$ and $Q$ are given by the following:
\eqn\gammaQX{\eqalign{
& \gamma_Q =  -{1\over2}\left(1-{1\over N_c^2}\right)\hat g+{1\over2N_c^2}\hat\lambda + \CO\left(\hat g^2,{\hat g\hat\lambda\over N_c^2},{\hat\lambda^2\over N_c^2}\right) \ ,\cr
& \gamma_X = {3\over 2(1+\epsilon)}\hat\lambda + \CO\left({\hat g\hat\lambda\over N_c^2},{\hat\lambda^2\over N_c^2}\right) \ .
}}
where we have introduced a large $N$ coupling for $\lambda$, $ \hat\lambda\equiv {N_c^2\,a^2\,\lambda^2\over8\pi^2}$. The beta functions consist of \nsvz, together with
\eqn\betadeformed{
\beta_{\hat\lambda} = 2\hat\lambda(2\gamma_Q+\gamma_X) \ .
}
While these equations are exactly solved by
\eqn\gammaQXsol{
\gamma_Q = -{\epsilon\over2},\quad \gamma_X=\epsilon \ , 
}
the couplings at the fixed point are determined perturbatively using \gammaQX,
\eqn\fpcplgl{\eqalign{
& \hat g_{*} = \hat g_{*,0} + {2\over 3N_c^2}\Big(\epsilon+\CO(\epsilon^2)\Big) + \CO\Big({1\over N_c^4}\Big)\cr
& \hat\lambda_* = {2\over3}\epsilon(1+\epsilon)+ \CO\left(\epsilon^2\over N_c^2\right) \ .
}}
At the new fixed point, we note that $\hat g_*$ differs from $\hat g_{*,0}$ at $\CO(\epsilon/N_c^2)$.

Given the couplings, we find the matrix of anomalous dimensions is given by:
\eqn\GammaQX{\eqalign{
\Gamma = \pmatrix{ { \partial\beta_{\hat g}\over\partial {\hat g}} & { \partial\beta_{\hat g}\over\partial\hat \lambda}\cr { \partial\beta_{\hat\lambda}\over\partial \hat g} & { \partial\beta_{\hat \lambda}\over\partial {\hat \lambda}}}\Bigg|_{\hat g=\hat g_{*},\hat\lambda=\hat\lambda_*}  
 &= \pmatrix{ -6\hat g_*^2 (\hat g_{*})\partial_{\hat g}\gamma_Q(\hat g_{*},\hat\lambda_*) & -6\hat g_*^2 f(\hat g_{*})\partial_{\hat\lambda}\gamma_Q(\hat g_{*},\hat\lambda_*)\cr
2\hat\lambda_* \partial_{\hat g}(2\gamma_Q+\gamma_X)(\hat g_{*},\hat\lambda_*) & 2\hat\lambda_*\partial_{\hat\lambda}(2\gamma_Q+\gamma_X)(\hat g_{*},\hat\lambda_*)}\cr
&=\pmatrix{\left(1+{4\over3N_c^2}\right)\nu_L + \CO({\epsilon^3/ N_c^2}) & -{3\epsilon^2\over N_c^2}+\CO(\epsilon^3/N_c^2)\cr
-{4\over3}\epsilon + \CO(\epsilon^2) +\CO(\epsilon/N_c^2)& 2\epsilon + {4\epsilon\over3N_c^2} + \CO(\epsilon^2/N_c^2) }
}}
The eigenvectors of this matrix correspond to scaling operators formed from $L$ and $X^{\dagger}X$, with anomalous dimensions given by the eigenvalues:
\eqn\eigenbzs{
2\epsilon  + {4\epsilon+\dots\over 3N_c^2} + \CO\Big({1\over N_c^4}\Big),\qquad 
\nu_L + {2\epsilon^2+\dots\over N_c^2} + \CO\Big({1\over N_c^4}\Big)\dots \ .
}
In \GammaQX\ and \eigenbzs, we have explicitly included all the terms that are reliably computed with one-loop anomalous dimensions. At the same time, we should emphasize that the $\nu_L$ appearing in \GammaQX\ and \eigenbzs\ is the {\it exact} anomalous dimension of $L$ in the undeformed BZ theory, i.e.\ including all higher loop corrections. It is important to capture these corrections even though they cannot be computed explicitly, so that the deformation only shifts $\nu_L$ by $\CO(1/N_c^2)$.  

We can arrive at the same results using the general formalism from section 2. According to \Gammaprime, we should have
\eqn\betasbzs{\eqalign{
\Gamma= \pmatrix{ \nu_L & 2c_L \epsilon\cr
 2c_L\epsilon & 2\epsilon} + \dots
 }}
so the eigenvalues are:
\eqn\betasbzsii{
2\epsilon + {4\epsilon^2c_L^2\over 2\epsilon-\nu_L} + \dots,\qquad \nu_L-{4\epsilon^2 c_L^2\over 2\epsilon-\nu_L} + \dots
}
If we plug in $c_L = \sqrt{2\over N_fN_c} = \sqrt{2(1+\epsilon)\over 3N_c^2}$, then we find perfect agreement with the correction to the $2\epsilon$ eigenvalue in \eigenbzs.  On the other hand, our $1/N_c^2$ correction to the $\nu_L$ eigenvalue does not match. This is because we are keeping enough terms in conformal perturbation theory to determine the correction to $\Delta_{X^\dagger X}$, but not enough terms for the correction to $\Delta_L$. For instance, a term $\sim c_i y_i\lambda^2$ in $\beta_{y_i}$ would correct $\nu_L$ at the same order as what one gets from mixing (but it doesn't affect the correction to $2\epsilon$ to leading order). While it would be interesting to also get these terms right, it is not necessary for our purposes and is beyond the scope of this paper.

Of course, given \deltaW, we know that  this example does not come close to testing whether the dimension of $X^\dagger X$ can be larger than $2 \Delta_X$; in terms of the classification of section 2.3, we are in the $\nu\ll\epsilon$ regime of ``scenario 1." This occurs in most BZ-like fixed points -- the $X^\dagger X$ OPE always contains an operator associated with an anomalous $U(1)$, like $L$, whose anomalous dimension $\nu_L$ typically obeys $\nu_L \ll \epsilon$. As we will show in the next subsection, it is possible to construct a model with all $\nu_L>2\epsilon$, but at the cost of significant tuning of parameters.

\centerline{\epsfxsize=0.8\hsize\epsfbox{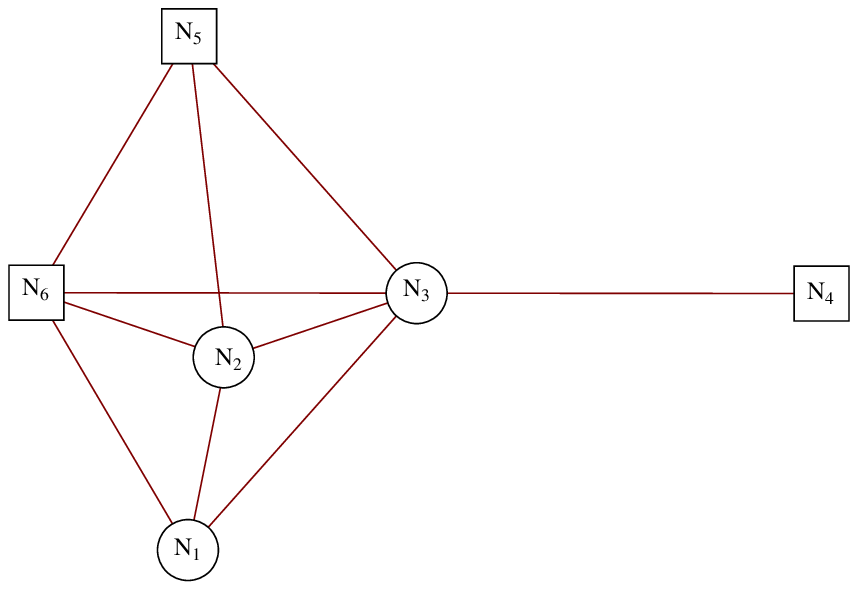}}
\noindent{\ninepoint\sl \baselineskip=8pt {\bf Figure 1}:{\sl $\;$
Quiver diagram for a viable generalized BZ model, where it is possible to achieve $\nu_L\gg \epsilon$ for all $L$. Circles denote $SU(N)$ gauge groups; squares denote $SU(N)_L\times SU(N)_R$ global symmetries. Links between nodes denote vector-like bifundamental fields. Finally, every triangle corresponds to a gauge and global-symmetry invariant Yukawa coupling. }}

\subsec{Generalized Banks-Zaks Fixed Points}

Clearly, to get a meaningful example with all $\nu_L> 2\epsilon$, we need to consider generalized BZ fixed points which have multiple small parameters and the freedom to dial $\epsilon\to 0$ without simultaneously dialing  any of the $\nu_L\to 0$. 

Shown in fig.\ 1 is a quiver diagram for a viable generalized BZ model. Although the model is quite complicated, it is the simplest example we have been able to find. The nodes in fig.\ 1 are labelled by 
$N_i$, $i=1,\dots,6$; circles indicate $SU(N_i)$ gauge groups, and boxes indicate $SU(N_i)_L\times SU(N_i)_R$ flavor symmetries.  Links between nodes (there are 10 altogether) denote vector-like bifundamental fields; we will label these fields by the pair of numbers corresponding to the nodes that they connect, e.g.\ the bifundamentals between nodes $N_1$ and $N_2$ will be called $Q_{12}$ and $ Q_{21}$. Our convention will be that the field is always a fundamental under the first node  and an anti-fundamental under the second node. Finally, every triangle corresponds to two gauge and global-symmetry invariant Yukawa couplings with different orientations. For instance, the 1-2-3 triangle corresponds to $\delta W = \lambda_{123} {\rm Tr}\,Q_{12}Q_{23}Q_{31} +\lambda_{123}' {\rm Tr}\, Q_{21} Q_{32} Q_{13}$. 

The non-anomalous, unbroken flavor symmetry of this model is $\prod_{i=4}^6 SU(N_i)_L\times SU(N_i)_R$, together with 5 baryonic symmetries left unbroken by the Yukawa couplings. In addition, at the fixed point there is an unbroken  ${\Bbb Z}_2$ symmetry under which $Q_{ij}\leftrightarrow  Q_{ji}$. We will choose to preserve this ${\Bbb Z}_2$ symmetry along the entire flow, to avoid unnecessary clutter. As a result, $\gamma_{ij}=\gamma_{ji}$ and $\lambda_{ijk}=\lambda_{kji}$.  One can also include the ${\Bbb Z}_2$ odd couplings and verify this explicitly.

 As we will show, in this model $\epsilon\equiv -2 \gamma_{12}$ can be dialed to zero while keeping all the gauge and Yukawa couplings (and hence all the $\nu_L$) nonzero. Therefore, we will couple $X$ to 
\eqn\OdefgenBZ{
\CO \sim {\rm Tr}\, Q_{12} Q_{21} \ .
} 
Before we launch into a more detailed analysis, let us discuss some general points:

\item{1.} $\CO$ appears to be the unique operator in this model whose anomalous dimension can be dialed to zero independently of the couplings.

\item{2.} Unlike in the previous example, $\CO$ is a singlet under the gauge and global symmetry of the quiver theory. The analysis of section 2 will apply directly, without having to deal with any of the complications discussed in appendix B. 

\item{3.} In general, there are a number of approximately dimension two scalar operators $L_i$ that could appear in the OPE of $X^\dagger$ and $X$. Clearly, any such operator must be a gauge and global singlet, and they must be invariant under $Q_{ij}\leftrightarrow Q_{ji}$. The complete list of compatible dimension two operators is
\eqn\LijgenBZ{
L_{ij} = {4\pi^2\over\sqrt{2N_iN_j}}{\rm Tr}\, (Q_{ij}^\dagger Q_{ij} + Q_{ji}^\dagger Q_{ji})\ .
} 
There are 10 such operators, and they are all explicitly broken by Konishi anomalies and the Yukawa couplings.  There are an additional ten, ${\Bbb Z}_2$-odd operators of dimension near two, of which five are the conserved baryonic currents.  These twenty operators can be combined to form the twenty $U(1)$ currents of the free theory that commute with the non-abelian flavor symmetry ($Q_{ij} \to e^{i \theta_{ij}} Q_{ij}$).  

\item{4.} For the choice of \OdefgenBZ, only one operator, $L_{12}$, appears in the $X^\dagger X$ OPE to leading order. Its leading order OPE coefficient is:
\eqn\conetwo{
c_{12} = \sqrt{\frac{2}{N_1 N_2} \ .}
}

\item{5.} This model has both Yukawa couplings and multiple gauge groups. Both are necessary conditions -- otherwise sending $\gamma_{ij}\to 0$ necessarily forces $\nu_{L}\to 0$ faster for some $L$.

\medskip

Now let us embark on a more detailed analysis of this model. We will follow the notation and conventions established in the previous subsection, with some minor additions. First, because of the complexity of this model, we will not be as careful about the double expansion in $N$ and $\epsilon$ as we were previously; we will stick to leading-order one-loop anomalous dimensions at large $N$. Second, since here we have many $N_i\to \infty$, we will  introduce an auxiliary parameter $N$, and define
\eqn\yepsdef{
x_i = {N_i\over N}, \quad \epsilon_i = {b_i\over N}, \quad \hat g_i = {N g_i^2\over 8\pi^2},\quad \hat \lambda_{ijk} = {N \lambda_{ijk}^2\over 8\pi^2}
}
where $b_i=3N_i-\sum_j N_j$  are the coefficients of the one-loop beta functions.  We will take $N\to \infty$ holding fixed $x_i$ and $\epsilon_i$.

For a general quiver of this type, the gauge beta functions are (we neglect the denominator in the NSVZ formula, because it will not matter for our discussion):
\eqn\betarew{\eqalign{
& \beta_{\hat g_i} = -\hat g_i^2(\epsilon_i +2 \sum x_j \gamma_{ij}) \ , \cr
& \beta_{\hat \lambda_{ijk}} =2 \hat \lambda_{ijk}(\gamma_{ij}+\gamma_{jk}+\gamma_{ki}) \ , \cr
& \gamma_{ij} =  -{1\over2}x_i \hat g_i - {1\over2}x_j \hat g_j + {1\over2}\sum_k x_k \hat \lambda_{ijk}  \ .
}}
Here $\hat g_i=0$ if $i$ is not a gauge node, and $\hat \lambda_{ijk}=0$ if $(ijk)$ is not a triangle. One can check that all of the anomalous dimensions $\gamma_{ij}$ (and hence all of the couplings $\hat g_i$, $\hat\lambda_{ijk}$) are fully determined by setting the beta functions to zero. 

In \betarew, the gauge and Yukawa interactions contribute with opposite signs to $\gamma_{ij}$. This is ultimately what allows us to tune $\gamma_{12}\to 0$ while keeping all the couplings nonzero (and real and positive). To demonstrate this explicitly, consider taking
\eqn\yansatz{
x_1 = {7\over4}+{3\over2}{\epsilon\over2-\epsilon}+\xi,\quad x_2 = 2+{4\over5}\xi,\quad x_3={9\over4}+{9\over5}\xi,\quad x_4=x_5=x_6=1 \ .
}
with $\epsilon,\,\xi\ll 1$. The origin of coupling-space corresponds to  $\epsilon=\xi=0$. Solving \betarew, one finds 
\eqn\gammaonetwo{
\gamma_{12}= -{1\over2}\epsilon
}
independent of $\xi$. Thus we can dial $\gamma_{12}\to 0$ while maintaining $\xi$, and hence the couplings, nonzero (and real and positive).
Correspondingly, at $\epsilon=0$, the dimensions of the $L_{ij}$ operators are all nonzero:
\eqn\Ldims{
\Delta_L \approx (18.99\xi,\, 7.57\xi,\, 6.38\xi,\, 2.78\xi,\, 2.18\xi,\, 0.37\xi,\, 8.10\xi^2,\, 2.25\xi^2,\, 0.08\xi^2) + \dots \ .
}
with $\dots$ denoting higher powers of $\xi$. 

To complete the construction of the model, we couple $X$ to $a\Tr\,Q_{12}Q_{21}$ via a Yukawa coupling $\lambda$. This introduces the beta function for $\hat\lambda\equiv {N^2a^2\lambda^2\over 8\pi^2}$ and modifies the anomalous dimension of $Q_{12}$, as in the previous subsection. Along with \betarew, the RG flow is now described by 
\eqn\betalambdagenBZ{\eqalign{
 & \beta_{\hat \lambda} =2\hat \lambda(2\gamma_{12}+\gamma_X)\cr
 & \gamma_X = {1\over2}x_1x_2\hat\lambda\cr
 & \gamma_{12} \to \gamma_{12} +{1\over 2N^2}\hat\lambda \ .
 }}
At the new fixed point, we again have $\gamma_X=-2\gamma_{12}=\epsilon$.
 
So as we dial $\epsilon\to 0$ while keeping the couplings nonzero, we will reach a regime where $\gamma_X=\epsilon \ll \nu_{L_{ij}} \ll 1$. This in turn provides a concrete example of the nontrivial part of ``scenario 1" from section 2.3, where mixing with the $L$ operators gives the dominant contribution to $\delta_{min}$ and forces it to be negative.   
We can see this explicitly by using the formulas \betalambdagenBZ\  to calculate the correction to the $X^\dagger X$ anomalous dimension in the limit \yansatz. For $\epsilon \ll \xi^2$, we find: 
\eqn\XXadimgenBZ{
\delta_{min} =\Delta_{min}-2\Delta_X \approx -{ 0.57\epsilon^2\over N^2 \xi^2} \ .
}
This is the parametric behavior of $\delta_{min}\sim -c_i^2\epsilon^2/\nu_i$ expected from Section 2, with the smallest $\nu_i\sim \xi^2$ dominating from \Ldims, and $c_i\sim 1/N$ according to \conetwo. We will leave a more detailed matching with conformal perturbation theory to Appendix C.

In summary, we have constructed an example that realizes all the regimes of ``scenario 1" of our conformal perturbation theory calculation.  This example serves two purposes.  First, it demonstrates that there is no fundamental obstacle to varying $\epsilon$ and $\nu_i$ independently.  Second, it illustrates how the bound derived in Section 2 arises in an actual example. 

\subsec{ Comments on double-trace operators}

In our treatment of the generalized BZ theories, we have neglected an important subtlety: the presence of double-trace operators.  In large $N$ theories,  the OPE of single trace operators will contain double trace operators with $\CO(1)$ OPE coefficients. 

When all the OPE coefficients are comparable in size, we are justified in focusing on the operators of the lowest dimension, as they will have the largest mixing with $X^{\dagger} X$.  In the weakly-coupled BZ theories, we determined the dimension of $X^{\dagger} X$ by isolating the mixing with single trace operators of approximate dimension 2, such as the Konishi currents. However, the OPE coefficients that control the mixing with these operators are suppressed by $\CO(1/N)$. Therefore, when $N \to \infty$, single trace operators do not mix with $X^{\dagger} X$ and do not alter its scaling dimension.  One may worry that the dominant effect arises from the double-trace operator $L_4= \CO^\dagger\CO$ which has approximate dimension 4 but an {\it unsuppressed} OPE coefficient with $\CO$. Since $1/N$ is the smallest parameter in the model, if $L_4$ has any unsuppressed mixing with $X^\dagger X$, then it will dominate the anomalous dimension calculation, despite being of much higher dimension

Fortunately, the double-trace operator is also decoupled from $X^\dagger X$ to leading-order in $1/N$. We see this from the calculation in appendix A: the $y\lambda^2$ term in $\gamma_X$ vanishes when $\Delta_L=2\Delta_\CO$. Deviations from this are proportional to $\Delta_L-2\Delta_\CO\sim 1/N^2$ for the double-trace operator in question. The double-trace operator then contributes at the same order as any other real scalar operator with $\nu\sim\CO(1)$, and its mixing with $X^\dagger X$ is subleading relative the Konishi currents which have $\nu\ll 1$. 

What about at higher orders in conformal perturbation theory? These correspond to higher order (four points and higher) correlation functions involving $\CO$, $\CO^\dagger$, and $L_4$.  In the $N \to \infty$ limit, we may use $L_4= \CO^\dagger\CO$ to find that non-vanishing correlation functions are disconnected and do not require additional counterterms.  New counter terms arise from connected correlation functions of $\CO$ and $\CO^{\dagger}$ and are necessarily $1/N$ suppressed, as standard $1/N$ power-counting shows. We conclude that the decoupling of the double-trace operators is robust to all orders in perturbation theory.

\newsec{Conclusions}

A very interesting question, with numerous potential phenomenological and formal applications, is:  what are the bounds on the lowest dimension operator that can appear in the OPE of scalar primaries? To date, most work on this question has focused on the constraints of unitarity and crossing symmetry \refs{\RattazziGJ \CaraccioloBX\RattazziYC\PolandWG \VichiUX{--}\PolandEY}, or on theories with gravity duals \FitzpatrickHH.  In this paper, we address how such bounds arise in perturbative SCFTs.  Using conformal perturbation theory, we found that a bound $\Delta_{X^{\dagger}X} < 2 \Delta_X$ can be seen directly from the beta functions.  

The calculations  in this paper were performed as a perturbative expansion in $\epsilon = \Delta_X - 1$.  In perturbative theories, the $\CO(\epsilon^2)$ contribution to $\Delta_{X^{\dagger} X}$ was computable because it is dominated by mixing with other operators with dimensions near two.  These results are  sufficient to cover a very broad class of perturbative SCFTs, including Banks-Zaks theories and their generalizations.  

Extending our results to all SCFTs would require a more comprehensive treatment of conformal perturbation theory in several regimes: 

\item{1.} We did not include all $\cO(\epsilon^2)$ contributions to the beta functions, including an $\CO(|\lambda|^4)$ contribution to the anomalous dimension of $X$.  These corrections will be relevant to strongly coupled theories.  

\item{2.} When the deformation breaks a global symmetry, the RG flow is more complicated and a complete analysis might be insightful.  In the appendix, we simply showed that the contributions to $\Delta_{X^{\dagger}X}$ from the broken currents are small in the cases of interest.  

\item{3.} We did not completely elucidate the limit where the dimension of a non-chiral operator approaches two.  As these operators become conserved currents in this limit, a description using {\it approximately} conserved currents would likely be more useful. 

\item{4.} Finally, as with all perturbative arguments, our results do not apply if contributions at a given order in perturbation theory are anomalously small.  A more systematic understanding of conformal perturbation theory would be helpful to address these and other questions.

\medskip

The calculations in this paper could be useful for understanding RG flows in other applications.  One advantage of our approach is that we have made all properties of the RG flow manifest.  The results in the paper are organized explicitly as a perturbative expansion in $\epsilon$ and are non-perturbative in any other small parameters present in explicit examples.  For this reason, our analysis could be applied to strongly coupled SCFTs that are weakly coupled to a free field.  For example, many theories contain a regime of parameters where the dimension of a gauge invariant chiral operator violates the unitarity bound and is believed to become a free field \refs{\KutasovIY{,}\BarnesJJ}. It is plausible that our results could describe these theories as the dimension of this operator approaches one.

\bigskip

\centerline{\bf Acknowledgements}

We would like to thank N.~Craig, T.~Dumitrescu, G.~Festuccia, T.~Hartman, S.~Knapen, Z.~Komargodski, D.~Poland and N.~Seiberg for helpful discussions.  The research of D.G is supported by the DOE under grant number DE-FG02-90ER40542 and the Martin A. and Helen Chooljian Membership at the Institute for Advanced Study. The research of D.S. partly supported by a DOE Early Career Research Award.

\appendix{A}
{Explicit Calculation of the Beta Functions}

\subsec{Setting up the calculation}

In this appendix, we will explicitly calculate the contributions to the beta functions quoted \gammaform\ and \betay.  We will largely follow the approach taken in \CraigRK.  However, we do offer an improved understanding of total derivative counterterms, and to our knowledge we perform the first ``two-loop" calculation in 4d conformal perturbation theory.

We would like to understand the renormalization group flow of the Langrangian \rgaction.  For convenience we will write the action as $S= S_{\cP_1} + \delta S$ with
\eqn\rgactionb{\eqalign{
\delta S &=\int d^4x\, d^4 \theta\, \left( \frac{1}{4\pi^2} (1+\delta Z_X) X^{\dagger} X(x,\theta,\bar\theta) +  \sum_i (y_i+\delta y_i) \Lambda^{-\nu_i} L_i(x,\theta,\bar\theta) \right)\cr
&\qquad +  \left( \int d^4z^+\, d^2 \theta \,\frac{1}{2\pi} \lambda\Lambda^\epsilon \cO X(z^+,\theta)  + \int d^4z^-\, d^2 \theta \,\frac{1}{2\pi} \lambda\Lambda^\epsilon \cO^\dagger X^\dagger(z^-,\bar\theta)\right) \ .
}}
As in the main text, we have taken $\Delta_{\cO} =  2 - \epsilon$ and $\Delta_{L_i} = 2 + \nu_i$. Here we are being careful to show explicitly the dependence on the superspace coordinates $(z^\mu)^\pm=x^\mu\pm i\theta\sigma^\mu\bar\theta$.  In order to achieve finite results, we will impose UV and IR cutoffs $\Lambda$ and $\ell^{-1}$ respectively.  We are free to choose a holomorphic renormalization scheme in which $\lambda$ only depends on the cutoff through its classical scaling, $\lambda = \lambda_0 (\Lambda/\Lambda_0)^{-\epsilon}$, where $\Lambda_0$ is some fixed scale. Also, by definition, $y_i$ refers to the part of the $L_i$ coupling which has the classical scaling, $y_i =y_{i0}(\Lambda/\Lambda_0)^{\nu_i}$. 

We will determine the beta functions by requiring that the correlation functions are independent of the cutoff scale $\Lambda$. The counterterms  $\delta Z_X$ and $\delta y_i$ in \rgactionb\ are fixed by this requirement. Let us write $\delta Z_X$ and $\delta y_i$ as an expansion in the couplings: 
\eqn\ZXexpand{\eqalign{
\delta Z_X &= a_{1}|\lambda|^2 + a_{1i} y_i + a_2 |\lambda|^4+a_{2i} |\lambda|^2 y_i + a_{2ij} y_i y_j + \dots\cr
\delta y_i&= b_{1i}|\lambda|^2 + b_{1ij} y_j + b_{2i} |\lambda|^4+b_{2ij} |\lambda|^2 y_j+ b_{2ijk} y_j y_k + \dots \ .\cr
}}
By expanding a general correlation function in powers of $\lambda$ and $y_i$ and performing the OPE repeatedly, we can fix the counterterms order by order.

For our calculations, we will need the $\CO\CO^\dagger$, $X X^\dagger$ two-point functions and the $L\CO\CO^\dagger$ three point function in superspace.  Superconformal invariance determine these up to overall normalization \OsbornQU. We fix the two-point function normalization to one, and then the three-point function normalization is the OPE coefficient:
\eqn\OOJ{\eqalign{
 & \langle\CO(z_1^+,\theta_1)\CO^\dagger(z_2^-,\bar\theta_2)\rangle = {1\over (X_{21}^+)^{2(2 - \epsilon)}},\qquad \langle X(z_1^+,\theta_1)X^\dagger(z_2^-,\bar\theta_2)\rangle = {1\over (X_{21}^+)^{2}}\cr
 & \langle\CO(z_1^+,\theta_1) \CO^\dagger(z_2^-,\bar\theta_2) L_i(x_3,\theta_3,\bar\theta_3)\rangle = {c_i  \over (X_{21}^+)^{2 - 2 \epsilon-\nu_i}(X_{23}^+)^{2+\nu_i}(X_{31}^+)^{2+\nu_i}} \ ,
}}
where 
\eqn\Xpdef{
(X_{ij}^+)^\mu \equiv  x_{ij}^\mu-i(\theta_i\sigma^\mu \bar\theta_j-\theta_j\sigma^\mu\bar\theta_i) - i\theta_{ij}\sigma^\mu\bar\theta_{ij} =  (z_i^-)^\mu -(z_j^+)^\mu + 2i\theta_j\sigma^\mu\bar\theta_i
}
is a supertranslation invariant interval. From \OOJ, we can read off some superspace OPEs that we will need for the calculation:
\eqn\OOXXOPE{\eqalign{
& \cO^{\dagger}(z_2^-,\bar\theta_2) \cO(z_1^+,\theta_1) \to  {1\over (X_{21}^+)^{2(2-\epsilon)}} + {c_i \over (X_{21}^+)^{2-2\epsilon - \nu_i}}  L_i(x_1,\theta_1,\bar\theta_2)+\dots\cr
& X^\dagger(z_2^-,\bar\theta_2)X(z_1^+,\theta_1) \to {1\over (X_{21}^+)^2} + X^\dagger X(x_1,\theta_1,\bar\theta_2) + \dots
}}
and
\eqn\LOXOPE{\eqalign{
 & L_i(x_3,\theta_3,\bar\theta_3)\CO(z_1^+,\theta_1) \to {c_i\over (X_{31}^+)^{2+\nu_i}} \CO(z_1^+,\theta_1) + \dots\cr
 & L_i(x_3,\theta_3,\bar\theta_3)\CO^\dagger(z_2^-,\bar\theta_2) \to {c_i\over (X_{23}^+)^{2+\nu_i}} \CO^\dagger(z_2^-,\bar\theta_2) + \dots\cr
}}
To be precise, the OPE limit  we are taking  in \OOXXOPE\ is $x_2\to x_1$, $\theta_2\to\theta_1$, $\bar\theta_1\to\bar\theta_2$. Here $\dots$ denote superconformal descendants, i.e.\ derivatives in position and in superspace acting on the superfield primary. These will not matter for any of our beta function calculations in the next subsection.

Finally, one comment about our regularization procedure. As mentioned above, we will impose UV and IR cutoffs of $1/\Lambda$ and $\ell$ respectively. These will be imposed as a hard point-splitting cutoff between any two operators. It is crucial for our calculations that this point-splitting cutoff be implemented in a supersymmetric way. We have found that the easiest way to do this is to work in superspace instead of components, and to be careful to always impose the cutoff on supertranslation invariant intervals $X_{ij}^+$ defined in \Xpdef. Other ways of imposing the cutoff, for instance on position differences $x_{ij}$, appear to break supersymmetry, leading to non-supersymmetric counterterms.

\subsec{Calculating the counterterms}

Now we are finally ready to calculate the counterterms. Let us imagine expanding a general correlation function in powers of $y_i$ and $\lambda$.

\medskip
{\bf 1.} At $\CO(y)$, we find:
\eqn\correxpandy{
\Big\langle \dots \int d^4x\,d^4\theta\Big( {1\over 4\pi^2}a_{1i}y_i\,   X^\dagger X+
 (y_i+b_{1ij}y_j )\, \Lambda^{-\nu_i} L_i  
 \Big) \Big\rangle \ .
}
The $y_i$ term is cutoff independent, and the other terms must be absent, since they are cutoff-dependent. So we conclude that
\eqn\abonei{
a_{1i}= b_{1ij}=0 \ .
}
Note that we are not allowed to use the equations of motion of the free theory to set $X^\dagger X$ to zero, since $X^\dagger X$ appears in the correlation function as an integrated operator, and the equations of motion can fail due to contact terms when operators coincide.

\medskip

{\bf 2.} At $\CO(|\lambda|^2)$, we find (now and henceforth dropping the $\langle \dots \rangle$):
\eqn\correxpandll{\eqalign{
& \int d^4x\, d^4\theta \, \Big( {1\over4\pi^2}a_{1}|\lambda|^2    X^\dagger X(x,\theta,\bar\theta) + b_{1i}|\lambda|^2 \,\Lambda^{-\nu_i}  L_i(x,\theta,\bar\theta) 
\Big) \cr
&\qquad +{1\over4\pi^2}|\lambda|^2 \Lambda^{2\epsilon} \int d^4z_1^+ \,d^4z_2^- d^2\theta_1 d^2\bar\theta_2\,  \CO X(z_1^+,\theta_1) \CO^\dagger X^\dagger (z_2^-,\bar\theta_2)
} }
The $a_{1}$ and $b_{1i}$ counterterms are required in order to cancel off the cutoff dependence in the third term when the integrated operators fuse to give $ X^\dagger X$ and $L_i$. To see this, we use the OPE \OOXXOPE\ on the third term:
\eqn\correxpandllii{\eqalign{
& \int d^4z_1^+ \,d^4z_2^-\, d^2\theta_1 \, d^2\bar\theta_2\, \CO X(z_1^+,\theta_1) \CO^\dagger X^\dagger (z_2^-,\bar\theta_2) = \cr
& \int d^4z_1^+ \,d^4z_2^-\, d^2\theta_1 \, d^2\bar\theta_2\,
\left( {1\over (X_{21}^+)^{2(2-\epsilon)}}X^\dagger X(x_1,\theta_1,\bar\theta_2) + {1\over (X_{21}^+)^{4-2\epsilon - \nu_i}} c_i L_i(x_1,\theta_1,\bar\theta_2)+\dots\right)
}}
Then we are free to change integration variables $z_2^-\to X_{21}^+$ and $z_1^+\to x_1$ with trivial Jacobian, so that \correxpandllii\ becomes
\eqn\correxpandlliii{
\int d^4X_{21}^+\, {1\over (X_{21}^+)^{2(2-\epsilon)}} \times \int d^4x\, d^4\theta\, X^\dagger X(x,\theta,\bar\theta) + \int d^4X_{21}^+\, {c_i\over (X_{21}^+)^{4-2\epsilon-\nu_i}} \times \int d^4x\,d^4\theta \, L(x,\theta,\bar\theta)
}
Performing the $X_{21}^+$ integrals with a short-distance cutoff $1/\Lambda$, plugging this back into \correxpandll, and requiring that the counterterms in \correxpandll\  cancel off the result, we finally obtain 
\eqn\aboneii{
a_{1}= {\pi^2\over\epsilon} ,\qquad  b_{1i}=  {c_i\over2(2\epsilon+\nu_i)} \ .
}
The appearance of the $\epsilon^{-1}$ in the counter-terms here is no different than the $(4-d)^{-1}$ that appears in counter-terms in dimensional regularization.   Although the counter-term will diverge in the $\epsilon \to 0$ limit, the $\epsilon^{-1}$ does not appear in the beta functions and RG flow is well behaved.

\medskip

{\bf 3.} At $\CO(y\lambda)$, we find (taking into account \abonei):
\eqn\correxpandyl{\eqalign{
 &{1\over2\pi}y_i\lambda\Lambda^{\epsilon-\nu_i} \int d^4x_3d^4\theta_3 \,d^4z_1^+ d^2\theta_1 \, L_i(x_3,\theta_3,\bar\theta_3) \CO X(z_1^+,\theta_1)  + c.c.\cr
} }
These terms must be cutoff independent, because in our holomorphic scheme, there are no counterterms which can cancel them. Indeed, when $L_i$ approaches $\CO X$ or $\CO^\dagger X^\dagger$, we can use the OPEs in \LOXOPE. Substituting these into \correxpandyl\ yields
\eqn\correxpandylope{\eqalign{
 &{1\over2\pi}y_i\lambda\Lambda^{\epsilon-\nu_i} \int d^4x_3d^4\theta_3 \,d^4z_1^+ d^2\theta_1 \, {c_i\over (X_{31}^+)^{2+\nu_i}} \CO X(z_1^+,\theta_1)  + c.c.\cr
} }
We are free to change integration variables from $x_3\to X_{31}^+$ and $\theta_3\to \theta_{31}$; then \correxpandylope\ is annihilated by the $d^4\theta_{31}$ integral. This change of variables also ensures that all the superconformal descendants in the $L\CO$ OPE do not contribute to renormalization of $\CO X$ in the superpotential.

\medskip

{\bf 4.} At $ \CO(y|\lambda|^2)$, we find:
\eqn\correxpandyll{\eqalign{
&  {1\over4\pi^2} y_i|\lambda|^2 \Lambda^{2\epsilon-\nu_i}\Bigg[ \int d^4z_1^+\, d^2\theta_1\, d^4z_2^-\,d^2\bar\theta_2\, d^4x_3\, d^4\theta_3\, L_i(x_3,\theta_3,\bar\theta_3) \CO X(z_1^+,\theta_1)\CO^\dagger X^\dagger (z_2^-,\bar\theta_2) \cr
&  + \Lambda^{-2\epsilon}\int d^4x_1\,d^4\theta_1 \,d^4x_2\, d^4\theta_2\, L_i(x_1,\theta_1,\bar\theta_1) \left(a_{1} X^\dagger X(x_2,\theta_2,\bar\theta_2)+ 4\pi^2 b_{1i} \Lambda^{-\nu_i} L_i(x_2,\theta_2,\bar\theta_2) \right)\cr
& + \Lambda^{\nu_i-2\epsilon}\int d^4x \,d^4\theta\, \left(a_{2i} X^\dagger X(x,\theta,\bar\theta) + 4\pi^2 b_{2ij}\Lambda^{-\nu_j}L_j(x,\theta,\bar\theta)\right)\Bigg]
 }}
The first term gives rise to cutoff dependence that must be cancelled by the counterterms.  We already fixed $a_{1}$ at a lower order in perturbation theory.  Clearly, $a_{1}$ does not play an important role -- $L$ and $X^\dagger X$ have a trivial OPE in the unperturbed theory and $a_1$ will cancel the contribution to this operator from the first term. The $b_{1i}$ and $b_{2ij}$  terms are relevant for the renomalization of  $y_k L_k$ and other non-chiral operators of $\cP_1$, but these contributions are negligible for our present purposes. What we are interested in is the $X^\dagger X$ counterterm represented by $a_{2i}$ that is required to cancel off the cutoff dependence when $L_i\CO\CO^\dagger\to1$ in the first term of \correxpandyll. In more detail, using the OPEs \OOXXOPE,  we have from the first term:
\eqn\correxpandyllii{\eqalign{
  &  \int d^4z_1^+\, d^2\theta_1\, d^4z_2^-\,d^2\bar\theta_2\, d^4x_3\, d^4\theta_3\, L_i(x_3,\theta_3,\bar\theta_3) \CO X(z_1^+,\theta_1)\CO^\dagger X^\dagger (z_2^-,\bar\theta_2) \cr
   &\quad \to \int d^4z_1^+\,d^2\theta_1 d^2\bar\theta_2\,  X^\dagger X(x_1,\theta_1,\bar\theta_2)\cr
   & \qquad \qquad\times \left( \int d^4z_2^-\,d^4x_3\,d^4\theta_3 \, \langle L_i(x_3,\theta_3,\bar\theta_3) \CO(z_1^+,\theta_1) \CO^\dagger(z_2^-,\bar\theta_2)\rangle  \right) \cr
   }}
Comparing with \correxpandyll, we see that the required counterterm is\foot{Actually, we are dropping here a potential contribution from $\int d^4x_1d^4\theta_1\,  |\lambda|^2 L_i X^\dagger X \times \int d^4x_2d^4\theta_2\, y_i L_i$. This vanishes when $\nu_i\to 0$ or when $\Delta_{L_i}=2\Delta_\CO$, the two special cases we are interested in.} 
\eqn\twoloopct{
   a_{2i} = -\Lambda^{2\epsilon-\nu_i}   \int d^4z_2^-\,d^4x_3\,d^4\theta_3 \, \langle L_i(x_3,\theta_3,\bar\theta_3) \CO(z_1^+,\theta_1) \CO^\dagger(z_2^-,\bar\theta_2)\rangle
   }
where the integrals should be performed at fixed $z_1^+$, $\theta_1$, and $\bar\theta_2$.   Using \OOJ\ to evaluate the three-point function, and changing integration variables from $(z_2^-,\,x_3,\,\theta_3,\bar\theta_3)\to (X_{23}^+,X_{31}^+,\theta_{31},\bar\theta_{31})$ (with trivial Jacobian), this becomes
\eqn\twoloopctii{ \eqalign{
 a_{2i} &= -\Lambda^{2\epsilon-\nu_i}  \int d^4X_{23}^+\,d^4X_{31}^+\,d^4\theta_{31}\,{c_i\over (X_{23}^+ + X_{31}^+ + 2i\theta_{31}\sigma\bar\theta_{31} )^{2-2\epsilon-\nu_i}(X_{23}^+)^{2+\nu_i}(X_{31}^+)^{2+\nu_i}}\cr
  &=c_i  (2\epsilon+\nu_i)(2-2\epsilon-\nu_i) \Lambda^{2\epsilon-\nu_i} \int { d^4X_{23}^+\,d^4X_{31}^+\over (X_{23}^+ + X_{31}^+ )^{4-2\epsilon-\nu_i}(X_{23}^+)^{2+\nu_i}(X_{31}^+)^{2+\nu_i}}\cr
  &\equiv {8\pi^4 c_i\over\nu_i-2\epsilon}\CI(\nu_i,\epsilon) 
 }}
where we have defined the dimensionless integral function
\eqn\Isimp{\eqalign{
\CI(\nu_i,\epsilon) & = {(\nu_i-2\epsilon)(2\epsilon+\nu_i)(2-2\epsilon-\nu_i)\over 8\pi^4}\Lambda^{2\epsilon-\nu_i}  \int { d^4X_{23}^+\,d^4X_{31}^+\over (X_{23}^+ + X_{31}^+ )^{4-2\epsilon-\nu_i}(X_{23}^+)^{2+\nu_i}(X_{31}^+)^{2+\nu_i}}\cr
}}
In general, this is a complicated function of $\nu_i$ and $\epsilon$, because we must take care to regularize the integral so that $|X_{23}^+|>1/\Lambda$, $|X_{31}^+|>1/\Lambda$, and $|X_{23}^++X_{31}^+|>1/\Lambda$. In two special cases, things simplify: 

\item{a.} If $\nu_i \ll 1$, then the singularities at $X_{23}^+=0$, $X_{31}^+=0$, and $X_{23}^++X_{31}^+=0$ are separately integrable, so the integral is dominated by the common singularity at $X_{23}^+,\,X_{31}^+\to 0$. We should be able to calculate this dominant part by doing the $X_{23}^+$ integral without regard to the $X_{23}^+=0$ and $X_{23}^+=-X_{31}^+$ singularities, and then doing the $X_{31}^+$ integral with a $1/\Lambda$ cutoff.  Doing so, we obtain $\CI=1$ plus higher orders in $\nu_i$ and $\epsilon$. 

\item{b.} If $\Delta_{L_i}=2\Delta_\CO$ (i.e.\ $\nu_i=2-2\epsilon$), as is the case for the double-trace operators discussed in section 3.3, then \Isimp\ simply vanishes. Therefore, in this case,  the wavefunction renormalization of $X$ is independent of $y_i$ to this order.

\subsec{Calculating the beta functions}

With the counterterms \ZXexpand\ in hand, it is straightforward to compute the beta functions for the couplings, keeping in mind we are interested in the {\it physical} couplings,
\eqn\physcoup{
\lambda^{phys} = \lambda Z_X^{-1/2},\qquad y_{i}^{phys} = y_i+\delta y_i \ .
}
Using the counter terms and classical running of the bare couplings, we find:
\eqn\betaphys{\eqalign{
 & \beta_{\lambda^{phys}} = -\epsilon\lambda^{phys} + \gamma_X \lambda^{phys}\cr
 &  \beta_{y_i^{phys}} = \nu_i y_i + \delta y_i ' = \nu_i y_i^{phys} -{1\over2} c_i |\lambda|^2 + \CO(y^2,\lambda^4,y\lambda^2)
 }}
where $\gamma_X \equiv -{1\over2}{d \log Z_X \over d\log\Lambda}$. Substituting \ZXexpand, \abonei, \aboneii, and \twoloopctii, we obtain for the anomalous dimension of $X$:
\eqn\gammaXgen{
\gamma_X = (\pi^2-4\pi^4\CI(\nu_i,2\epsilon)c_i y_i)|\lambda|^2 + \CO(\lambda^4,\lambda^2 y^2)
}
At this order in perturbation theory, we are free to substitute $\lambda^{phys}$, $y_i^{phys}$ for $\lambda$, $y_i$ in \betaphys\ and \gammaXgen. This completes our derivation of the beta functions used in the body of the paper.

 
 \appendix{B}{Generalizing to the case with global symmetry}
 
In this appendix, we will generalize the discussion in Section 2 to situations where $\cO$ transforms in some nontrivial representation ${\bf r}$ of the global symmetry group $G$ of $\cP_1$.  

By definition, when $\cO$ is charged, the appropriate conserved current multiplets $J^a$ will appear in the $\cO$-$\cO^{\dagger}$ OPE:  
\eqn\OopeJ{
\cO_m(x) \cO_{\bar n}^{\dagger} (0) = g_{m \bar n} |x|^{-2 \Delta_{\cO}} + \sum_a |x|^{2 - 2 \Delta_{\cO} } T^a_{m \bar n} J^a + \sum_{i,p} |x|^{\Delta_L - 2 \Delta_{\cO} } (c_i)_{m\bar n p} (L_{i})_p+\ldots\ .}  
Here $m$, $n$ and $p$ are global symmetry indices.  $T^a_{m\bar n}$ are the generators of $G$ in the representation ${\bf r}$; their normalization is fixed by the requirement that the $J^a$'s are CFT-canonically normalized. We are free to choose $g_{m\bar n} = \delta_{m\bar n}$ at $\cP_1$. \foot{The meaning of this ``metric" becomes more clear when $\cP_1$ is a point on a conformal manifold.  In that case, $g_{m \bar n}$ defines a metric on that space and can have non-trivial dependence on the exactly marginal couplings \KutasovXB.}  In writing \OopeJ , we have allowed for the possibility that $L_i$ transforms in some representation ${\bf h}$ of $G$.  Allowing for charged $\CO$ and $L_i$ complicates the situation considerably.

As before, we will deform the theory by the superpotential $W \supset \frac{1}{2\pi} \lambda \cO X$, where now $\lambda$ transforms in the ${\bf \bar r}$ representation of $G$ (the global symmetry indices on $\lambda$ and $\CO$ are implicit).  This deformation breaks the global symmetry, allowing the currents to acquire anomalous dimensions and mix with $J^X \equiv X^\dagger X$ and $L_i$.    

In order to arrive at a non-trivial bound, we should forbid the appearance of conserved currents in the OPE of $X$ and $X^{\dagger}$ at $\cP_2$.  That is, we will demand that our theory does not contain an unbroken $U(1)$ current formed from a linear combination of $J^X$ and $J^a$. These satisfy
\eqn\currents{\eqalign{
& \bar D^2 J^a 
= \lambda T^a \CO X + \ldots \ ,\cr 
& \bar D^2 J^X = \lambda \cO X +\ldots \ .
}}
Therefore $\lambda T^a$ and $\lambda$ should be linearly independent vectors.
Stated more formally, we require that the only real solution to $D^2( \sum_a C_a J^a + C_X J^X ) = 0$ is $C_a = C_X = 0$.  Using \currents, this requirement can rewritten (to leading order) as
\eqn\constraint{
\sum_a C_a   \lambda T^a  + C_X \lambda = 0 \quad \Rightarrow\quad C_a = C_X = 0  \ .}

After deforming the theory by $\lambda \cO X$, the theory will undergo an RG flow to the new fixed point $\cP_2$.  As before, we will work in a holomorphic basis where the only renormalization is of operators in the Kahler potential.  We should therefore consider the following action
\eqn\rgactionb{\eqalign{
\cL = \cL_{\cP_1}&+ \int d^4 \theta \left( \frac{1}{4 \pi^2} Z_X J^X + \sum_a Z_a J^a +  \sum_i \Lambda^{2-\Delta_{i}} y_i\cdot  L_i + \ldots  \right) \cr &+  \left[ \int d^2 \theta \frac{1}{2\pi} \Lambda^{\epsilon}\lambda \cO X  + {\rm h. c.} \right] \ .}}
At one loop, $\delta Z_a = - 2 \pi^2 \lambda T^a \lambda^\dagger \, {\rm log}(\Lambda) $  \GreenDA.

The beta function for $y_i$ is calculated in the same way as before:
\eqn\betayb{
\beta_{y_i} = \nu_{i} y_i - \frac{1}{2} \lambda c_i\lambda^\dagger +  \CO(\lambda^4,y\lambda^2,y^2)\ .
}
The beta function for $\lambda$ is not useful in the holomorphic basis, as the fixed point appears at $\lambda \to \infty$.  To study the fixed point, we will move to a non-holomorphic basis by absorbing $Z_a$ and $Z_X$ into the definition of $\lambda$.  This can be accomplished using \currents~to rewrite
\eqn\coupling{
\lambda \to \lambda - \frac{1}{2} Z_X \lambda - \frac{1}{2} Z_a \lambda T^a \ . }
The RG flow for $\lambda$ is governed by
\eqn\rgflowb{
\beta_\lambda  = \lambda ( - \epsilon + \gamma_X) + \gamma_a \lambda T^a \ , }
where $\gamma_a = - \frac{1}{2} \frac{\partial \log Z_a}{\partial \log \Lambda}$. We compute the contributions to $Z_X$ as before to find (at small $\nu_i$)
\eqn\gammaX{
\gamma_X = \pi^2\lambda \lambda^\dagger - 4 \pi^4  \lambda (y_i\cdot c_i)\lambda^\dagger +\CO(y^2\lambda^2,\lambda^4) \ . }
We will not compute all the contributions to $\gamma_a$, but we will simply note that it takes the form:
\eqn\gammaa{
\gamma_a =  \pi^2 \lambda  T^a  \lambda^\dagger+ \cO( \lambda^4, y\lambda^2,y^2) \ . 
}
From \constraint, we see that $\beta_\lambda = 0$ requires\foot{This conclusion is true, even when higher order terms in \currents~are included. The solutions to $\beta_\lambda = 0$ are equivalent to the solutions of $D^2( \sum_a \gamma_a J^a + (\gamma_X-\epsilon) J^X ) = 0$, so our conclusion follows from forbidding conserved currents. } that 
\eqn\gammaXafp{
\gamma_X = \epsilon,\qquad \gamma_a = 0\ .
}  
To leading order, this implies $ \lambda T^a \lambda^\dagger = 0$ for every $a$.

We now wish to compute the scaling operators around the fixed point as before.  In general, $\lambda$ is an $r$-dimensional vector of couplings that admits many independent deformations. We are only interested in those ``directions"  that correspond irrelevant operators, namely the broken currents $J^X$, $J^a$. Other deformations either correspond to exactly marginal operators or total derivatives. In order to focus on just the broken currents, let us specialize to the deformations around the fixed point which correspond to them via \currents:
\eqn\lambdadefc{
\lambda= \lambda_* +  \eta_0 \lambda_* +  \eta_a \lambda_*  T^a
}
with $\eta_0$, $ \eta_a$ real. In other words, we are trading the vector of couplings $\lambda$ for a subset parametrized by real $(\eta_0,\eta_a)$ which correspond to the broken currents. Note that the unbroken currents automatically drop out of this since they satisfy $\lambda_* T^a =0$. 

We would like to know the beta functions for $(\eta_0,\eta_a)$ to linear order. We can read them off from \rgflowb, using the linear independence of $\lambda_*$ and $\lambda_* T^a $ and the fact that $-\epsilon+\gamma_X,\,\gamma_a\sim \CO(\eta)$:
\eqn\betac{\eqalign{
& \beta_{\eta_0} = -\epsilon+\gamma_X + \CO(\eta^2) = -\epsilon+\pi^2\lambda\lambda^\dagger  - 4\pi^4  \lambda (c_i\cdot y_i)\lambda^\dagger +\CO(\lambda^4,y^2\lambda^2)+ \CO(\eta^2)\cr
& \beta_{\eta_a} = \gamma_a + \CO(\eta^2) = \pi^2\lambda T^a \lambda^\dagger  + \CO(\lambda^4,y \lambda^2,y^2)+\CO(\eta^2)
}}
Combining this and \betayb, we finally obtain the matrix of beta function derivatives:
\eqn\betacder{\eqalign{
\Gamma &=  \pmatrix{ {\partial \beta_{\eta_0}\over\partial\eta_0} &  {\partial \beta_{\eta_0}\over\partial\eta_a} &  {\partial \beta_{\eta_0}\over\partial y_i} \cr
{\partial \beta_{\eta_b}\over\partial\eta_0} &  {\partial \beta_{\eta_b}\over\partial\eta_a} &  {\partial \beta_{\eta_b}\over\partial y_i} \cr
{\partial \beta_{y_j}\over\partial\eta_0} &  {\partial \beta_{y_j}\over\partial\eta_a} &  {\partial \beta_{y_j}\over\partial y_i}}\Bigg|_{\eta=0,y_i=y_{i*}} \cr
& = \pmatrix{ 
2\epsilon & 0 & -4\pi^4\lambda c_i\lambda^\dagger \cr
0 & 2 \pi^2 \lambda \{T^b,\, T^a\} \lambda^\dagger & \CO(\lambda^2,y)\cr
- \lambda c_i \lambda^\dagger & \CO(\lambda^2) & \nu_i} + \CO(\lambda^4, y \lambda^2,y^2)
}}
We can diagonalize the matrix $ 2 \pi^2 \lambda \{T^b,\, T^a\} \lambda^\dagger \to \tilde \nu^a \delta^{a b}$, and shift by $2\epsilon$, to write this as
\eqn\betadiagonal{
\Delta\Gamma =  \pmatrix{ 
0 & 0 & -4\pi^4\lambda c_i\lambda^\dagger \cr
0 & \tilde \nu^a-2\epsilon & \CO(\lambda^2,y)\cr
- \lambda c_i \lambda^\dagger & \CO(\lambda^2) & \nu_i-2\epsilon} + \CO(\lambda^4, y \lambda^2,y^2) \ .}
In a general model, the mixing of $J_X$ with $J_a$  is potentially important.  To get a feeling for this, let us simplify things somewhat by ignoring the distinction between $\lambda$ and $y$ and just taking $\lambda^2\sim y \sim \epsilon$. Also, let us drop the higher orders in \betadiagonal, as these are irrelevant for our discussion. Then the characteristic equation for the eigenvalues of \betadiagonal\ is
\eqn\betadiagonalce{
(\tilde\nu^a-2\epsilon-\delta)\Big((\nu_i-2\epsilon-\delta)\delta +4\pi^4(\lambda c_i\lambda^\dagger)^2 \Big) + \CO(\epsilon^2)\times \delta \approx 0
}
The last term in \betadiagonalce\ can be traced back to the $\CO(\lambda^2y)$ and $\CO(y^2)$ corrections to $\gamma_a$ which we have not computed. If $\tilde\nu^a$ is not close to $2\epsilon$, this term is higher order in $\epsilon$ and can be safely ignored, meaning the leading order for $\delta$ is as before. However, if $\tilde\nu^a=2\epsilon + \CO(\epsilon^2)$, then the last term in \betadiagonalce\ cannot be ignored for determining the leading order $\delta$.

To summarize so far: we have seen that it is valid to neglect the mixing with the $J_a$ operators, as long as their anomalous dimensions are not extremely degenerate with $2\epsilon$. If $\tilde\nu_a-2\epsilon\sim\CO(\epsilon^2)$, then the mixing with $J_a$ cannot be ignored, and more study (beyond the scope of this work) is required. 

We will now conclude our discussion by describing two important classes of models where the mixing can always be neglected, regardless of the $J_a$ anomalous dimensions.

\medskip

\lfm{\bf Case 1:} {\it $L_i$ are singlets under $G$.} The correlation functions of $L_i$ with $J^a$ are restricted by the representation of $L_i$ under $G$.  When all the $L_i$ are singlets of $G$, Ward identities require that $ \langle L_i L_j J^a \rangle = 0$ (or equivalently that $J^a$ is not in the OPE of $L_i$ and $L_j$).  As a result, the higher order correction to $\gamma_a$ take the form
\eqn\gammaasinglet{
\gamma_a =  \pi^2 \lambda T^a \lambda^\dagger + \cO( \lambda^4, y\lambda T^a \lambda^\dagger ) \ . 
}
Furthermore, $y_i$ is a flavor singlet and therefore
\eqn\betayb{
\beta_{y_i} = \nu_{i} y_i - \frac{1}{2} c_i \lambda \lambda^\dagger +  \CO(\lambda^4,y\lambda^2,y^2)\ .
}
with the $c_i$'s now just numbers, not matrices. As a result,
\eqn\betadiagonala{
 \Gamma =  \pmatrix{ 
2\epsilon & 0 & -4\pi^4c_i\lambda \lambda^\dagger \cr
0 & \tilde \nu^a & 0 \cr
-c_i  \lambda \lambda^\dagger & 0 & \nu_i} + \CO(\epsilon^2) \ .}
Here we have used $\lambda T^a \lambda^{\dagger} \sim \epsilon^2$.  Up to order $\epsilon^2$ corrections, $\tilde \nu_a$ are eigenvalues and do not mix with $J_X$ and $L_i$.  So after diagonalizing the $\tilde\nu^a$, $\nu_i$ block as above, one finds that the contribution from mixing with $\tilde\nu^a$ is subleading in $\lambda$, $y$, even if they are completely degenerate with $2\epsilon$. So the results of Section 2 are unchanged. 

\lfm{\bf Case 2:} {\it  Large $N$ limit.} In the Banks-Zaks examples we studied in Section 3, the OPE coefficients are controlled by a separate small parameter, $1/N$.  More generally, if the theory has a large $N$ parameter which controls the sizes of the $\CO^\dagger\CO$ OPE coefficients,
\eqn\OOlargeN{
\langle \CO^\dagger\CO L\rangle \sim \langle \CO^\dagger\CO J\rangle \sim 1/N
}
and such that other OPE coefficients are also suppressed by powers of $N$, specifically
\eqn\LLJ{
\langle L_i L_j J^a\rangle \sim 1/N^\alpha,\quad \alpha>0
}
then the conclusions of section 2 are robust. (This should include all generalized BZ theories and also theories with weakly-coupled gravity duals.) Using \OOlargeN\ and \LLJ, the large $N$ counting of \betadiagonal\ is
\eqn\betadiagonalN{
 \Gamma =  \pmatrix{ 
2\epsilon & 0 & \CO(1/N) \cr
0 & \tilde \nu^a & \CO(1/N^{1+\alpha})\cr
\CO(1/N) \ & \CO(1/N) & \nu_i} +\dots \ .}
The $\CO(1/N^{1+\alpha})$ came from two contributions to $\gamma_a$ -- the contribution at $\CO(y \lambda^2)$ involves $\CO\CO^\dagger L_i\to J^a$, which involves two OPE coefficients; and the contribution at $\CO(y^2)$ arises from $L L\to J$, plus the fact that $y\sim \CO(1/N)$ from \fixedcouplings. So we conclude that the correction from mixing with $\tilde \nu^a$ is suppressed by $1/N^{2+\alpha}$ and vanishes relative to the $\CO(1/N^2)$ contributions computed in Section 2.  As a result, the conclusions are unchanged.  

 \appendix{C}{More on the Generalized Banks-Zaks Example}

In this appendix, we would like to make contact between the analysis of section 2 and our generalized Banks-Zaks example, by rederiving \XXadimgenBZ\ using the general formula  \eigensolapproxii. 

In order to use the results of section 2, we must determine the operators in the OPE of $\CO \propto Q_{12}Q_{21}$ and $\CO^{\dagger}$ along with their OPE coefficients and dimensions.  This example is controlled by the large $N$ expansion and we will only be concerned with results at leading order in $1/N$.  As discussed in section 3, the only dimension approximate dimension two operator in the OPE is $\cO(x) \cO^{\dagger} (0)\to L_{12}$ where
\eqn\Lijgenonetwoagain{
L_{12} = {4\pi^2\over\sqrt{2N_1N_2}}{\rm Tr}\, (Q_{12}^\dagger Q_{12} + Q_{21}^\dagger Q_{21})\ ,
} 
with OPE coefficient
\eqn\conetwoagain{
c_{12} = \sqrt{\frac{2}{N_1 N_2} \ .}
}
The problem with directly applying \eigensolapproxii\ is that $L_{12}$ is not a scaling operator in the original undeformed model. So instead of \eigensolapproxii, we should have
\eqn\deltaminSb{
\delta_{min} = - 4\epsilon^2 c_{12}^2 (\Gamma^{-1})_{12;12} + \dots \ .
}
where $\Gamma$ is the matrix of anomalous dimensions in the $L_{ij}$ basis. The last remaining step is to find the transformation into this basis starting from the matrix of anomalous dimensions obtained from the beta functions \betarew. This change of basis  is implemented by the Konishi anomaly and $U(1)$ equations of motion. Adding $y_{ij} L_{ij}$ to the K\"ahler potential is equivalent to rotating $Q_{ij}$ and $Q_{ji}$ by the axial phase 
\eqn\Qijrot{
Q_{ij} \to e^{iq_{ij}y_{ij}}Q_{ij},\quad Q_{ji} \to e^{iq_{ij}y_{ij}}Q_{ji}
}
where the charge $q_{ij} \propto 1/\sqrt{x_ix_j}$ can be read off of the $\langle Q_{ij}^\dagger Q_{ij} L_{ij}\rangle$ three point function. This in turn induces a change in the superpotential and gauge couplings, via the current non-conservation equation (for a nice discussion of this, see \CachazoRY)
\eqn\currentnoncon{
\bar D^2 L_{ij} = q_{ij} \left( {\partial W\over\partial Q_{ij}} Q_{ij}+{\partial W\over\partial Q_{ji}} Q_{ji} + {N_j\over 16\pi^2} {\rm Tr}\, W_{\alpha i}^2+ {N_i\over 16\pi^2} {\rm Tr}\, W_{\alpha j}^2\right)
}
The third and fourth terms correspond to the Konishi anomaly terms, and they are present only if the $i$th or $j$th node is gauged, respectively.  Finally, from \currentnoncon, we read off the shift in the couplings (our conventions for the gauge kinetic term are the standard ones, in which $\CL\supset {1\over 4g^2}\int d^2\theta\, {\rm Tr}\,W_\alpha^2$)
\eqn\currentnonconii{
\delta \CL =\sum_{i>j} \int d^4\theta\, y_{ij} L_{ij}\quad \to\quad \delta \hat \lambda_{ijk} =2(y_{ij} q_{ij}+y_{ik}q_{ik}+y_{jk}q_{jk}) \hat \lambda_{ijk},\quad \delta \hat g_i = -2\sum_j y_{ij} q_{ij} x_j \hat g_i^2
} 
So this is a linear map $S$ which takes the couplings $y_{ij}$ to the couplings $(\hat g,\,\hat\lambda)$. If we then compute $\Gamma$ using the beta functions $\betarew$, we obtain it in the $L_{ij}$ basis by acting with $\Gamma\to S^{-1}\Gamma S$.  Substituting this into \deltaminSb, we find perfect agreement with \XXadimgenBZ.

\listrefs
\bye